\documentclass[12pt]{article}
\usepackage{amssymb}
\usepackage{epsf}
\usepackage{lscape}
\usepackage[cp866]{inputenc}
\usepackage[T2A]{fontenc}
\usepackage[english]{babel}
\usepackage{amsmath}
\begin{document}
\title{\Large\bf {Determination of the $ ^3{\rm {He}}+\alpha \to\rm {^7Be}$ asymptotic
normalization coefficients (nuclear vertex constants) and their
application for extrapolation of the $^3{\rm
{He}}(\alpha,\gamma)^7{\rm {Be}}$ astrophysical $S$  factors
 to  the solar energy region }}
\author{S.B. Igamov, Q.I. Tursunmahatov  and R. Yarmukhamedov \thanks{Corresponding author, E-mail: rakhim@inp.uz}}
\maketitle {\it {Institute of Nuclear Physics, Uzbekistan Academy of
Sciences,100214 Tashkent, Uzbekistan}}

 \begin{abstract}
 A new analysis of the modern precise measured astrophysical $S$  factors
for the direct capture $^3He(\alpha,\gamma)^7{\rm {Be}}$ reaction
[B.S. Nara Singh {\it et al.}, Phys.Rev.Lett. {\bf 93},  262503
(2004); D. Bemmerer {\it et al.}, Phys.Rev.Lett. {\bf 97},  122502
(2006); F.Confortola {\it et al.}, Phys.Rev.C {\bf 75}, 065803
(2007), T.A.D.Brown {\it et al.}, Phys.Rev. C {\bf 76},  055801
(2007) and A Di Leva, {\it et al.},Phys.Rev.Lett. {\bf 102}, 232502
(2009)] populating to the ground and first excited states of $^7{\rm
Be}$ is carried out based on the modified two - body potential
approach.    New estimates are obtained for the $^{\glqq}$indirectly
determined\grqq\, values of the asymptotic normalization constants
(the nuclear vertex constants) for $^3{\rm {He}}+\alpha\to{\rm
{^7Be}}$(g.s.)  and $^3{\rm {He}}+\alpha\to{\rm {^7Be}}$(0.429 MeV)
as well as the astrophysical $S$  factors $S_{34}(E)$ at E$\le$ 90
keV, including $E$=0. The  values of asymptotic normalization
constants have been used for getting information about the
$\alpha$-particle spectroscopic factors for the mirror
(${\rm{^7Li}}{\rm {^7Be}}$)-pair.
\end{abstract}


\section{Introduction}

\hspace{0.7cm}  The  $^3{\rm {He}}(\alpha,\gamma)^7{\rm {Be}}$
reaction is one of the  critical links in the $^7{\rm {Be}}$ and
$^8B$ branches of the $pp$--chain of solar hydrogen burning [1--3].
The total capture rate determined by processes of this chain is
sensitive to the cross section $\sigma_{34}(E)$ (or the
astrophysical $S$  factor $S_{34}(E)$ ) for the $^3{\rm
{He}}(\alpha,\gamma)^7{\rm {Be}}$ reaction and predicted neutrino
rate varies as $[S_{34}(0)]^{0.8}$  \cite{Bah82,Bah92}.

Despite the impressive improvements in our understanding of the
$^3{\rm {He}}(\alpha,\gamma)^7{\rm {Be}}$ reaction  made in the past
decades (see Refs [4--11] and references therein), however, some
ambiguities connected with both the extrapolation of the measured
cross sections for the aforesaid reaction to the solar energy region
and the theoretical predictions for $\sigma_{34}(E)$ (or
$S_{34}(E)$) still exist and they may  influence the predictions of
the standard solar model \cite{Bah82,Bah92} .

Experimentally, there are two types of data for the $^3{\rm
{He}}(\alpha,\gamma)^7{\rm {Be}}$ reaction at extremely low
energies: i) six measurements based on detecting of $\gamma$-rays
capture  (see \cite{Adel98} and references therein) from which the
astrophysical $S$  factor $S_{34}(0)$ extracted by the authors of
those works changes within the range 0.47$\le S_{34}(0)\le$0.58
${\rm {keV\,\,\, b}}$ and  ii) six measurements based on detecting
of $^7{\rm {Be}}$ (see \cite{Adel98} references therein as well as
[6-11]) from which $S_{34}(0)$ extracted by the authors of these
works changes within the range 0.53$\le S_{34}\le$0.63 ${\rm
{keV\,\,\,b}}$.  All of these measured data have a similar energy
dependence for the astrophysical $S$ factors $S_{34}(E)$.
Nevertheless, the adaptation  of the available energy dependencies
predicted in \cite{Kaj86,PD} for the extrapolation of each of the
measured data to low experimentally inaccessible energy regions,
including $E$=0, leads to  a value of $S_{34}(0)$ that differs from
others and this difference exceeds the experimental uncertainty.

The theoretical calculations of   $S_{34}(0)$ performed within
different methods also show  considerable spread  [12,14--19] and
the result depends on a specific model used. For example, the
resonating-group method calculations of $S_{34}(0)$ performed in
Ref.\cite{Kaj86} show considerable sensitivity to the form of the
effective NN interaction used and the estimates have been obtained
within the range of 0.312$\le S_{34}(0)\le$ 0.841 ${\rm
{keV\,\,\,b}}$.

The estimation of $S_{34}(0)$=0.52$\pm$ 0.03 ${\rm {keV\,\,\,b}}$
\cite{Igam97} also should be  noted.   The latter has been obtained
from the analysis of the experimental astrophysical $S$ factors
\cite{Osb82}, which were  performed within the framework  of the
standard two-body potential model in the assumption  that the
dominant contribution to the peripheral reaction comes from the
surface and external regions of the nucleus $^7{\rm {Be}}$
\cite{Chris61}. At this, in \cite{Igam97} the contribution from the
nuclear interior ($r< r_{cut}$, $r_{cut}$=4 fm) to the amplitude is
ignored. In this case, the astrophysical $S$ factor is directly
expressed in terms  of the nuclear vertex constants (NVC) for the
virtual decays ${\rm {^7Be}}\to\alpha +^3{\rm {He}}$ (or respective
the asymptotic normalization coefficients (ANC) for ${\rm
{^3He}}+\alpha\to{\rm {^7Be}}$) \cite{Blok77,Blokh2008}. As a
result, in Ref. \cite{Igam97}, the NVC-values   for the virtual
decays $ {\rm {^7Be}}{\rm {(g.s.)}}\to\alpha +{\rm {^3He}}$ and
${\rm {^7Be}}{\rm {(0.429\,\, MeV)}}\to\alpha +{\rm {^3He}}$ were
obtained, which were then used for calculations of the astrophysical
$S$ factors   at $E<$180 keV, including $E$=0.   However, the values
of the ANCs (or NVCs) ${\rm {^3He}}+\alpha\to{\rm {^7Be}}$ and the
$S_{34}(0)$ obtained in \cite{Igam97} may not be enough accurate
associated both with the aforesaid assumption in respect to  the
contribution from the nuclear interior ($r< r_{cut}$) and with a
presence of the spread in the experimental data \cite{Osb82} used
for the analysis. As far available values of these ANCs obtained in
\cite{PD,Nol01}, they depend noticeably on a specific model used.
Therefore, determination of precise experimental values of the ANCs
for ${\rm {^3He}}+\alpha\to{\rm {^7Be}}$(g.s.) and ${\rm
{^3He}}+\alpha\to{\rm {^7Be}}$(0.429 MeV) is highly desirable since
it has direct effects in the correct extrapolation of the ${\rm
{^3He}}(\alpha,\gamma){\rm {^7Be}}$ astrophysical $S$ factor at
solar energies.

Recently, a modified two-body potential approach (MTBPA) was
proposed in \cite{Igam07} for the peripheral direct capture ${\rm
A}(a,\gamma){\rm B}$ reaction, which is based on the idea proposed
in paper \cite{Chris61}   that low-energy direct radiative captures
of particle $a$ by light nuclei ${\rm A} $ proceed mainly in regions
well outside the range of the internuclear $a{\rm A}$ interactions.
One notes that in MTBPA  the direct astrophysical $S$ factor is
expressed in terms of ANC for ${\rm A}+a\to {\rm B}$ rather than
through the spectroscopic factor for the nucleus ${\rm B}$ in the
(${\rm A}+a$) configuration as it is made within the standard
two-body potential method \cite{Robert81,Kim87}. In
Refs.\cite{Igam07,Igam08,Art2009}, MTBPA   has been successfully
applied to the radiative proton and $\alpha$-particle capture by
some light nuclei. Therefore, it is  of great interest to apply the
  MTBPA for   analysis of the $ {\rm
{^3He}}(\alpha,\gamma){\rm {^7Be}}$ reaction.

In this work      new analysis of the modern precise experimental
astrophysical $S$  factors for the direct capture $^3{\rm
{He}}(\alpha,\gamma)^7{\rm {Be}}$ reaction at extremely low energies
($\gtrsim$ 90 keV)  [6-11] is performed within the MTBPA
\cite{Igam07} to obtain $^{\glqq}$indirectly determined\grqq\,
values both of the ANCs (the NVCs) for $^3{\rm {He}}+\alpha\to{\rm
{^7Be}}$(g.s.) and $^3{\rm {He}}+\alpha\to{\rm {^7Be}}$(0.429 MeV),
and of $S_{34}(E)$ at $E\le$ 90 keV, including $E$=0.   In this work
we quantitatively show that the $^3{\rm {He}}(\alpha,\gamma)^7{\rm
{Be}}$ reaction within the aforesaid energy region is mainly
peripheral and one can extract ANCs  for $^3{\rm {He}}+\alpha\to
{\rm {^7Be}}$ directly from the $ ^3{\rm {He}}(\alpha,\gamma){\rm
{^7Be}}$ reaction where the ambiguities inherent for the standard
two -body potential model calculation of the $^3{\rm
{He}}(\alpha,\gamma){\rm {^7Be}}$ reaction, which  is connected with
the choice of the geometric parameters (the radius $R$ and the
diffuseness $a$) for the Woods--Saxon potential and the
spectroscopic factors \cite{Moh93,Dub95}, can be reduced in the
physically acceptable limit, being within the experimental errors
for the $S_{34}(E)$.

The contents of this paper are as follows. In Section 2  the results
of  the analysis of the precise measured astrophysical $S$  factors
for the direct radiative capture $^3{\rm {He}}(\alpha,\gamma)^7{\rm
{Be}}$ reaction is presented (Subsections 2.1--2.3). The conclusion
is given in Section 3. In Appendix  basic formulae of the modified
two-body potential approach to the direct radiative capture $^3{\rm
{He}}(\alpha,\gamma)^7{\rm {Be}}$ reaction are given.

 \section{Analysis of   $^3{\rm {He}}(\alpha,\gamma)^7{\rm {Be}}$ reaction}

\hspace{0.6cm} Let us write    $l_f$ ($j_f$) for the relative
orbital (total) angular moment  of $^3{\rm {He}}$ and
$\alpha$-particle in nucleus $^7{\rm {Be}}\,\,(\alpha+{\rm
{^3He}})$, $l_ i$ ($j_ i$) for the orbital (total) angular moment
of the relative motion of the colliding particles in the initial
state. For the ${\rm {^3He}}(\alpha,\gamma){\rm {^7Be}}$ reaction
populating to the ground and first excited ($E^*$=0.429 MeV;
$J^{\pi}$=1/2$^-$) states of ${\rm {^7Be}}$, the values of $j_f$ are
taken to be equal to 3/2 and 1/2, respectively, the value of $l_f$
is taken to be equal to 1 as well as   $l_i$=0, 2 for the
$E1$-transition and $l_ i$=1 for the $E2$-transition.

The basic formulae used for the analysis are presented in Appendix.

\subsection{The asymptotic normalization coefficients for
 $^3{\rm {He}}+\alpha\to^7{\rm {Be}}$}

\hspace{0.7cm}   To determine the ANC values for ${\rm
{^3He}}+\alpha\to{\rm {^7Be}}$(g.s)  and ${\rm {^3He}}+\alpha\to{\rm
{^7Be}}$(0.429 MeV) the recent experimental astrophysical $S$
factors, $S^{exp}_{l_fj_f}(E)$, for the ${\rm
{^3He}}(\alpha,\gamma){\rm {^7Be}}$   reaction populating to the
ground ($l_f$=1 and $j_f$=3/2) and first excited ($E^*$=0.429 $MeV$;
$J^{\pi }=1/2^-$,\,\,\,$l_f$=1 and $j_f$=1/2) states [6--11] are
reanalyzed based on the   relations (A1)--(A7).
 The experimental data analyzed by us   cover the energy
ranges $E$=92.9--168.9 keV [7--9], 420--951 keV \cite{Nara04},
327--1235 keV \cite{Brown07} and 701--1203 keV \cite{Di2009} for
which only the external capture is substantially dominant
\cite{Moh2009,Chris61}. Also,  in \cite{Con07} the experimental
astrophysical $S$ factors for the the ${\rm
{^3He}}(\alpha,\gamma){\rm {^7Be}}$ reaction populating to the first
and excited states of the ${\rm {^7Be}}$ nucleus have been separated
only for the energies of $E$=92.9, 105.6 and 147.7 keV .    Whereas,
in \cite{Brown07} the experimental astrophysical $S$  factors have
been separated for all experimental points of $E$ from  the
aforesaid energy region   by means of detecting the prompt $\gamma$
ray (the prompt method) and by counting the ${\rm {^7Be}}$ activity
(the activation).

The Woods--Saxon potential split with a parity ($l$-dependence) for
the spin-orbital term proposed by the authors of Refs. [30--32] is
used here for the calculations  of both bound state  radial wave
function $\varphi _{l_fj_f}(r)$ and scattering wave function
$\psi_{l_ij_i}(r)$.  Such  the choice   is based on the following
considerations. Firstly, this potential form is justified from the
microscopic point of view because it makes it possible to take into
account the Pauli principle between nucleons in $^3{\rm {He}}$- and
$\alpha$-clusters in the ($\alpha+{\rm {^3He}}$) bound state by
means of inclusion of deeply bound states forbidden by the Pauli
exclusion principle.  The latter imitates  the additional node ($n$)
arising in the wave functions of $\alpha-{\rm {^3He}}$ relative
motion  in $^7{\rm {Be}}$ similarly as the result of   the
microscopic resonating-group method \cite{Kaj86}. Secondly, this
potential describes well the phase shifts for $\alpha{\rm
{^3He}}$-scattering in the wide energy range \cite{Neu75,Neu83} and
reproduces the energies of low-lying states of the $^7Be$ nucleus
\cite{Dub84}.

\begin{figure}[!h]
\epsfxsize=10.cm \centerline{\epsfbox{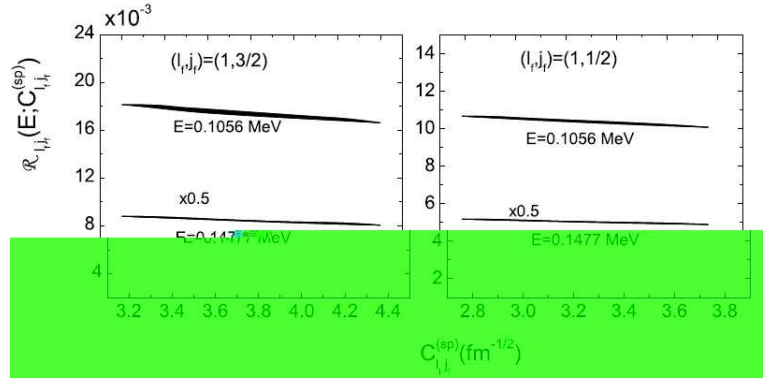}}
\caption{\label{fig1} The dependence of ${\cal
{R}}_{l_fj_f}(E,C^{(sp)}_{l_fj_f})$ as a function of the
single-particle ANC, $C^{(sp)}_{l_fj_f}$, for  the $^3{\rm
{He}}(\alpha,\gamma)^7{\rm {Be}}$(g.s.) ($(l_f,j_f)$=(1,3/2))  and
 $^3{\rm {He}}(\alpha,\gamma)^7{\rm {Be}}$(0.429
MeV  ($(l_f,j_f)$=(1,1/2)) reactions at  different energies $E$.}
\end{figure}

We vary the geometric parameters (radius $R$ and diffuseness $a$) of
the adopted Woods--Saxon potential in the physically acceptable
ranges ($R$ in 1.62--1.98 fm and $a$ in 0.63--0.77 fm \cite{Igam07})
in respect to the standard values ($R$=1.80 fm and $a$=0.70 fm
\cite{Neu75,Neu83}) using the procedure of the depth adjusted to fit
the binding energies. As an illustration,  Fig.\ref{fig1} shows
plots of the ${\cal {R}}_{l_fj_f}(E,C^{(sp)}_{l_fj_f})$  dependence
on the single-particle ANC, $C^{(sp)}_{l_fj_f}$  for $l_f$= 1 and
$j_f$=3/2 and 1/2 only for  the two  values of   energy  $E$. The
width of the band for these curves  is the result of the weak
$^{\glqq}$residual\grqq\,$(R,a)$-dependence of the ${\cal
{R}}_{l_fj_f}(E,C^{(sp)}_{l_fj_f})$  on the parameters $R$ and $a$
(up to $\pm 2\%$) for the
$C^{(sp)}_{l_fj_f}=C^{(sp)}_{l_fj_f}(R,a)=const$
\cite{Igam07,Gon82}. The same dependence is also observed at other
energies. For example, for fig.1 plotted for $E$=0.1056 (0.1477) MeV
overall uncertainty ($\Delta_{{\cal {R}}}$ ) of  the function ${\cal
{R}}_{l_fj_f}(E,C^{(sp)}_{l_fj_f})$ in respect to the central values
${\cal {R}}_{l_fj_f}(E,C^{(sp)}_{l_fj_f})$ corresponding to the
central values of  $C^{(sp)}_{l_fj_f}(1.80,0.70)$ comes to
$\Delta_{{\cal {R}}}$=$\pm$ 4.5($\pm$4.5)\%     for the ground state
of $^7{\rm {Be}}$ and $\Delta_{{\cal {R}}}$ =$\pm$ 3.4($\pm$2.9)\%
for the excited state of $^7{\rm {Be}}$. As it is seen from here
that the $^3{\rm {He}}(\alpha,\gamma)^7{\rm {Be}}$(0.429 MeV)
reaction is slightly more peripheral than the $^3{\rm
{He}}(\alpha,\gamma)^7{\rm {Be}}$(g.s.) reaction since the binging
energy  for  $^7{\rm {Be}}$(0.429 MeV) is less than that for $^7{\rm
{Be}}$(g.s). The similar dependence of ${\cal
{R}}_{l_fj_f}(E,C^{(sp)}_{l_fj_f})$ on the $C^{(sp)}_{l_fj_f}$
values  is observed for the other aforesaid  energies $E$ and the
value of $\Delta_{{\cal {R}}}$  is  no more than $\sim\pm$ 5.0\%. It
follows from here that the condition (A2) is satisfied for the
considered reaction at the    energies above mentioned within the
uncertainties not exceeding the experimental errors of
$S^{exp}_{l_fj_f}(E)$. It should be noted that values of
$\Delta_{{\cal {R}}}$ becomes larger as the energy $E$ increases
(for $E$ more than 1.3 MeV).

Thus, over the energy region 92.9$\le$E$\le$ 1200 keV the ${\cal
{R}}_{l_fj_f}(E,C_{l_fj_f}^{(sp)})$-dependence on
$C_{l_fj_f}^{(sp)}$  is exactly sufficiently weak being within the
experimental uncertainties for the $S^{exp}_{l_fj_f}(E)$. Such
dependence is apparently  associated also with the  indirect taking
into account the Pauli principle mentioned above  within the nuclear
interior in the adopted nuclear $\alpha{\rm {^3He}}$ potential
leading   as a whole to   reduction of the contribution from the
interior part of the radial matrix element into the ${\cal
{R}}_{l_fj_f}(E,C_{l_fj_f}^{(sp)})$ function, which is typical for
peripheral reactions.

\begin{figure}[!h]
\epsfxsize=10.cm \centerline{\epsfbox{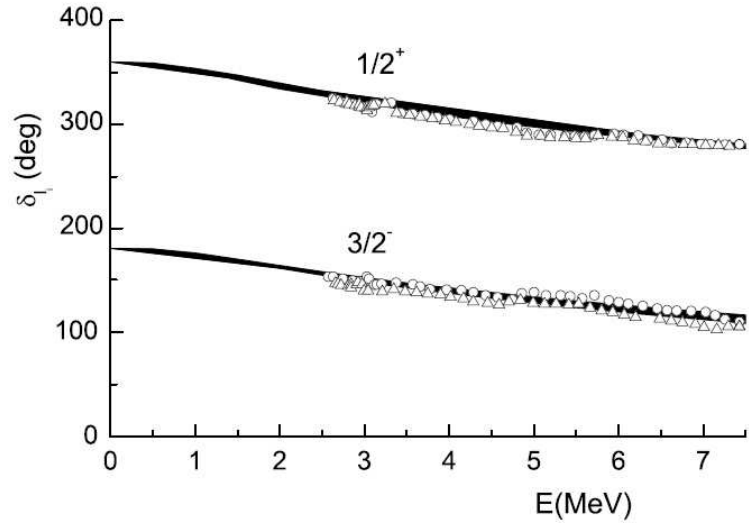}}
\caption{\label{fig2}The energy dependence of the $\alpha^3{\rm
{He}}$-elastic scattering phase
 shifts for the $s_{1/2}$ and $p_{3/2}$   partial waves. The experimental
data are from  \cite{Barn64,ST1967}. The bands are our calculated
data. The width of the bands for fixed energies corresponds to the
variation of the parameters $R$ and $a$ of the adopted Woods--Saxon
potential within the intervals of $R$=1.62 to 1.98 fm and $a$=0.63
to  0.77 fm.}
\end{figure}

We also calculated  the $\alpha^3{\rm {He}}$-elastic scattering
phase shifts by variation of the parameters $R$ and $a$ in the same
range for the adopted Woods--Saxon potential. As an illustration,
the results of the calculations corresponding to the $s_{1/2}$ and
$p_{3/2}$ waves are only presented in Fig.\ref{fig2} in which the
width of the bands corresponds to a change of phase shifts values
with respect to variation of values of the $R$ and $a$ parameters.
As it is seen from Fig.\ref{fig2}, the experimental phase shifts
\cite{Barn64,ST1967} are well reproduced within uncertainty of about
$\pm$ 5\%. The same results   are also obtained for the $p_{1/2}$
and $d_{5/2}$ waves.

\begin{figure}[!h]
\epsfxsize=10.cm \centerline{\epsfbox{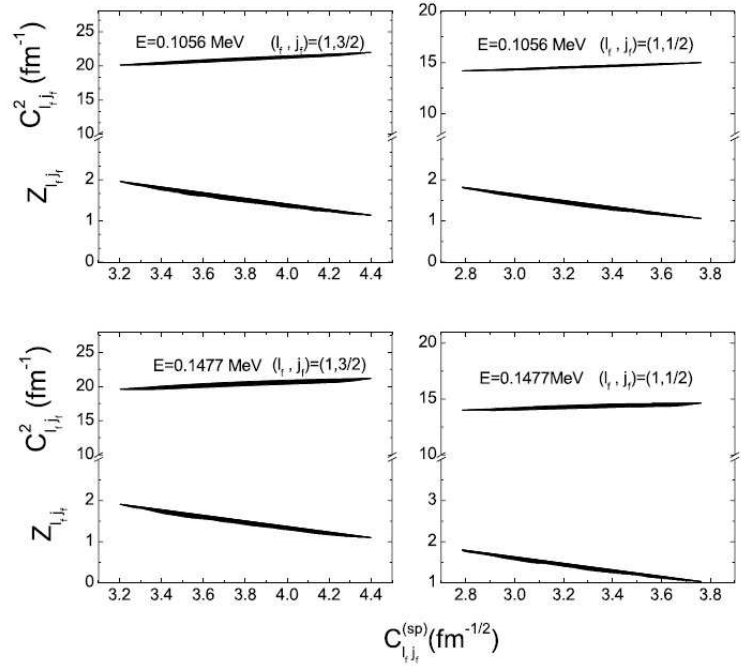}}
\caption{\label{fig3}The dependence of the ANCs $C_{l_fj_f}$ (upper
band) and the spectroscopic factors $Z_{l_fj_f}$ (lower band) on the
single-particle ANC $C^{(sp)}_{l_fj_f}$  for the $^3{\rm
{He}}(\alpha,\gamma)^7{\rm {Be}}$(g.s.) (the left column, $
(l_f,j_f)$=(1,3/2)) and $^3{\rm {He}}(\alpha,\gamma)^7{\rm
{Be}}$(0.429 MeV) (the right column, $(l_f,j_f)$=(1,1/2)) reactions
at  different energies $E$.}
\end{figure}

This circumstance allows us to test the condition (A3)  at the
energies of $E$= 92.9, 105.6 and 147.7 keV for which the $^3{\rm
{He}}(\alpha,\gamma)^7{\rm {Be}}$(g.s.) and $^3{\rm
{He}}(\alpha,\gamma)^7{\rm {Be}}$(0.429 MeV) astrophysical $S$
factors were separately measured  in  \cite{Con07}. As an
illustration,  for the same energies $E$  as in Fig.\ref{fig1} we
present in Fig.\ref{fig3} (the upper panels) the results of
$C_{l_fj_f}^2$-value calculation given by Eq.(A3)
($(l_f\,\,j_f)$=(1$\,\,$ 3/2) and (1$\,\,$ 1/2) ) in which instead
of the $S_{l_fj_f}(E)$ the experimental $S$ factors for the $^3{\rm
{He}}(\alpha,\gamma)^7{\rm {Be}} $ reaction populating to the ground
and first excited states of $^7{\rm {Be}}$ were taken. It should be
noted that the same dependence occurs for other considered energies.
The calculation shows the obtained $C_{l_fj_f}^2$ values also weakly
(up to 5.0 \%) depend  on the $C^{(sp)}_{l_fj_f}$-value.
Consequently,   the $^3{\rm {He}}(\alpha,\gamma)^7{\rm {Be}}$
reaction within the considered energy ranges is peripheral and  a
use of parametrization in terms of the ANCs given by Eq.(A1) is
adequate to the physics of the   reaction under consideration.
However, the values of the spectroscopic factors $Z_{1\,\,3/2}$ and
$Z_{1\,\,1/2} $ corresponding to the $(\alpha +^3{\rm
{He}})$-configuration for $^7{\rm {Be}}$(g.s.) and $^7{\rm
{Be}}$(0.429 keV), respectively, change  strongly about 1.7 times
 since   calculated $\tilde
{S}_{l_fj_f}(E)$ that vary by 1.75 times (see, the lower panels in
Fig.\ref{fig3}).

\begin{figure}[!h]
\epsfxsize=14.0 cm \centerline{\epsfbox{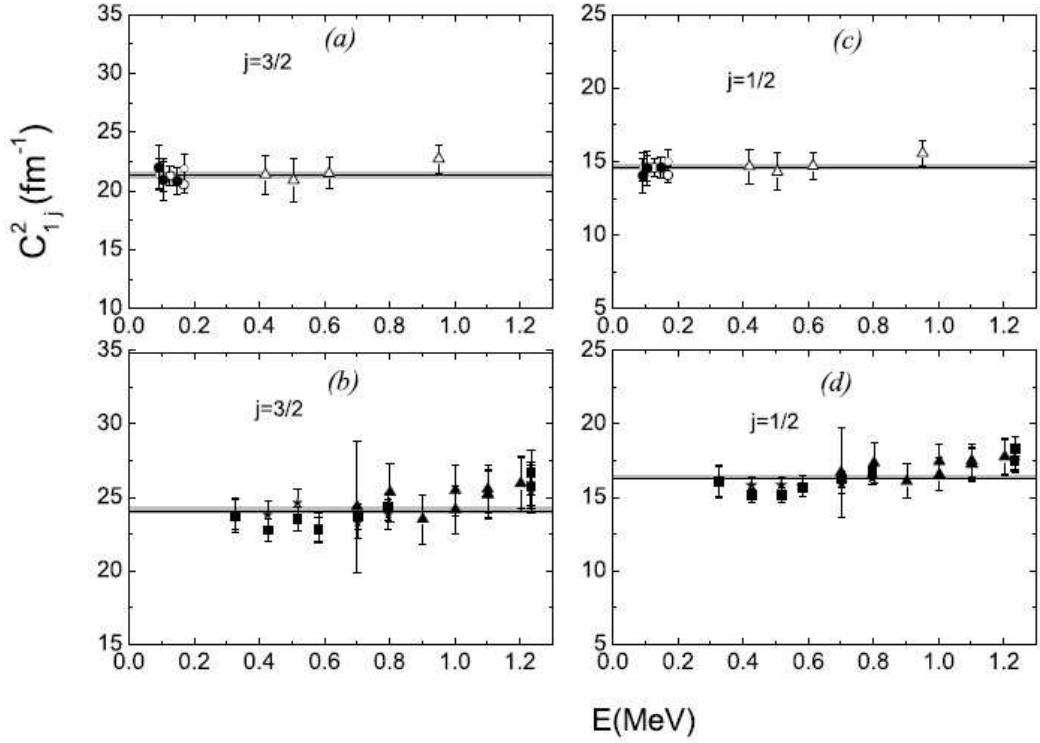}}
\caption{\label{fig5} The  values of the ANCs, $C^2_{1\,\,3/2}$ and
$C^2_{1\,\,3/2}$, for   $\alpha +{\rm {^3He}}\to{\rm {^7Be}}$(g.s.)
((\emph{a}) and (\emph{b})) and $\alpha +^3{\rm {He}}\to^7{\rm
{Be}}$(0.429 MeV) ((\emph{c}) and (\emph{d})) for each energy $E$
experimental point.     The open triangle  and cycle symbols  and
filled triangles (filled star(the activation), square (the prompt
method) symbols) are data  obtained by using the total (separated)
experimental astrophysical $S$ factors from  \cite{Nara04}(the
activation), \cite{Bem06,Con07} (the activation and the prompt
method) and \cite{Di2009}(the  ${\rm {^7Be}}$ recoils) (from
\cite{Brown07}), respectively, while filled circle symbols are data
obtained from the separated experimental astrophysical $S$ factors
from Refs.[7--9]. The solid lines present our results for the
weighted means. Everywhere the width of each of the band is the
weighted uncertainty.}
\end{figure}

For  each  energy $E$  experimental point ($E$=92.9, 105.6 and 147.7
keV) the values of the ANCs are obtained for   $\alpha + {\rm
{^3He}}\rightarrow{\rm {^7Be}}(g.s.)$ and $\alpha +{\rm
{^3He}}\rightarrow {\rm {^7Be}}$(0.429 MeV) by using the
corresponding experimental astrophysical $S$  factor
($S_{1\,\,3/2}^{exp}(E)$ and $S_{1\,\,1/2}^{exp}(E)$, (the
activation)) \cite{Bem06,Con07} in the ratio of the r.h.s. of the
relation (A1) instead of the $S_{l_fj_f}(E)$ and the central values
of ${\cal {R}}_{l_fj_f}(E,C^{(sp)}_{l_fj_f})$ corresponding to the
adopted values of the parameters $R$ and $a$.      The results of
the ANCs, $(C_{1\,\,3/2}^{exp})^2$ and $(C_{1\,\,1/2}^{exp})^2$, for
these three energy $E$ experimental points are displayed in
Fig.\ref{fig5}  $a$ and $c$ (filled circle symbols) and the second
and third columns of Table \ref{table1}. The uncertainties pointed
in this figure correspond to those found from (A1) (averaged square
errors (a.s.e.)), which include the total experimental errors
(a.s.e. from the statical and systematic uncertainties) in the
corresponding experimental astrophysical $S$ factor and the
aforesaid uncertainty in the ${\cal
{R}}_{l_fj_f}(E,C^{(sp)}_{l_fj_f})$. One should   note that the same
results for the ANCs  are obtained when $S^{exp}_{34}(E)$
($S^{exp}_{1\,3/2}(E)$ and $R^{exp}(E)$) \cite{Bem06,Con07}  are
used in Eq.(A5) (in Eq.(A6) and (A7)) instead of $S_{34}(E)$
($S_{1\,3/2}(E)$ and $R(E)$). Then in Eq.(A6), inserting the
averaged means of $\lambda_C$ ($\lambda_C$=0.666), obtained from the
three data,  and replacing of the $S _{34}(E)$ in the l.h.s. of
Eq.(A5) with $S^{exp}_{34}(E)$ for the others, the two experimental
points of energy $E$ ($E$=126.5 and 168.9 keV) from
\cite{Bem06,Gy07}, four one $E$ ($E$=420.0, 506.0, 615.0 and 951.0
keV) from \cite{Nara04},
  the three one $E$ ($E$=93.3, 106.1 and 170.1 keV) from
\cite{Con07} and the ten one $E$ ($E$=701--1203 keV) from
\cite{Di2009} can also determine values of ANCs, $C_{1\,\,3/2}^2$
and $C_{1\,\,1/2}^2$. The results of the ANCs $(C_{1\,j_f}^{exp})^2$
for   $\alpha +^3{\rm {He}}\rightarrow {\rm {^7Be}}$(g.s.) and
$\alpha +^3{\rm {He}}\rightarrow{\rm {^7Be}}$(0.429 MeV) are
displayed in Fig.\ref{fig5} in which  the open cycle and  triangle
symbols obtained from  the analysis of the data of [6--9] as well as
the filled triangles symbols obtained from the analysis of   the
data of \cite{Di2009}(the ${\rm {^7Be}}$ recoils). Besides, the
results obtained from the data of [7--9] are also presented in the
second and third columns of Table \ref{table1}. The same way the
values of the ANCs are obtained by using the separated experimental
astrophysical $S$ factors ($S_{1\,\,3/2}^{exp}$ and
$S_{1\,\,1/2}^{exp}$) \cite{Brown07}. The results for these ANCs are
also presented in Figs.\ref{fig5}$b$ and $d$  both for the
activation (filled star symbols) and for the prompt method (filled
square symbols).

As it is seen from Figs.\ref{fig5}, for each of the independent
measured experimental astrophysical  $S$ factors the ratio in the
r.h.s. of the relation (A4) does not practically depend on the
energy $E$ within the experimental uncertainties, although absolute
values of the corresponding experimental astrophysical $S$ factors
for the reactions under consideration depend noticeably on the
energy and change by up to about 1.7 times when $E$ changes from
92.6 keV to 1200 keV. This fact allows us to conclude that the
energy dependence of the experimental astrophysical $S$  factors
[6--11] is well determined by the calculated function ${\cal
{R}}_{l_fj_f}(E,C^{(sp)}_{l_fj_f})$ and ${{\cal
R}}_{13/2}(E,C_{13/2}^{(sp)})+\lambda_C{{\cal
R}}_{11/2}(E,C_{11/2}^{(sp)})$.   Hence, the experimental
astrophysical $S$  factors presented in [6--11] can be used as an
independent source of reliable information about the ANCs for
$\alpha +^3{\rm {He}}\rightarrow {\rm {^7Be}}$(g.s.) and $\alpha
+^3{\rm {He}}\rightarrow{\rm {^7Be}}$(0.429 MeV). Also, in
Fig.\ref{fig5} and   Table \ref{table2} the weighted means of the
ANCs-values and their uncertainties (the solid line and the band
width, respectively), derived both separately from each experimental
data and from all of the experimental points, are presented.

As it is seen also from the first and second (fifth and sixth) lines
of Table \ref{table2}  the weighted means of the ANCs-values for
$\alpha +^3{\rm {He}}\rightarrow {\rm {^7Be}}$(g.s.) and $\alpha
+^3{\rm {He}}\rightarrow{\rm {^7Be}}$(0.429 MeV) obtained by the
analysis performed separately for the activation, the prompt method
and the ${\rm {^7Be}}$ recoils of the experimental data from the
works [6--9] (\cite{Brown07,Di2009}) are in a good agreement with
one another.  These results are the first ones of the present work.
Nevertheless, the weighted means \cite{An99} of the ANC-values
obtained by using separately the experimental data of the works
Refs.[6--9]   and  of the works Refs.\cite{Brown07,Di2009}
noticeably differ from one another (up to 1.13 times for $\alpha
+^3{\rm {He}}\rightarrow {\rm {^7Be}}$(g.s.) and up to 1.12 times
for $\alpha +^3{\rm {He}}\rightarrow{\rm {^7Be}}$(0.429 MeV), see
the parenthetical figures in Table \ref{table2}). The main reason of
this difference is in the systematical discrepancy observed in
absolute values of the experimental astrophysical $S$ factors
measured by authors of works Refs [6--9](the set I) and of the works
Ref. \cite{Brown07,Di2009} (the set II, see Fig.\ref{fig6}). Also,
the central values of  the weighted means for the ANC-values for
$^3{\rm {He}}+\alpha \rightarrow{\rm {^7Be}}$(g.s.) and $^3{\rm
{He}}+\alpha\rightarrow {\rm {^7Be}}$(0.429 MeV) obtained from all
of the experimental data [6--11], which  is presented in the last
line of Table \ref{table2}, differ up to 10\% more (3\% less) than
those deduced from the data of [6--9](\cite{Brown07,Di2009}). As, at
present, there is no a reasonable argument to prosecute to some of
these experimental data measured by two groups ([6-9](the set I) and
\cite{Brown07,Di2009}(the set II)), it seems, it is reasonable to
obtain  the weighted means of the ANCs   derived  from all these
real experimental ANCs   with upper and lower  limits corresponding
to the experimental data of the set II and Set I, respectively. This
leads to the asymmetric uncertainty for the weighted means of ANCs
and this is caused with the aforesaid systematical discrepancy
observed in absolute values of the experimental data of the sets I
and  II (see the last line of Table \ref{table2}). In this
connection, from our point of view a new precisely measurement of
$S_{34}^{exp}(E)$ is highly encouraged since  it allows one  to get
an  additional information about the ANCs. Nevertheless, below we
will use these ANCs for extrapolation of the astrophysical $S$
factors at energies down, including $E$=0. The corresponding values
of NVCs obtained by using Eq.(A8) are given in Table \ref{table2}.

 A comparison of the present result    and
that obtained in paper \cite{Igam97} shows that  the underestimation
of the contribution both of the nuclear interior and of the nuclear
exterior indeed occurs in \cite{Igam97} since the contribution of
the nuclear interior ($r<$ 4.0 fm) to the calculated astrophysical
$S$ factors and use the experimental data [6-11] more accurate than
those in Ref.\cite{Igam97}  can   influence the extracted values of
the ANCs. Besides, one would also like to note  that in reality the
values of the ANCs, $C_{1\,\,3/2}$ and $C_{1\,\,1/2}$, should not be
equal, as it was assumed in \cite{PD} and  the values of
$C_{1\,\,3/2}^2=C_{1\,\,1/2}^2$=14.4 fm$^{-1}$ were obtained from
the analysis of $S_{34}^{exp}(E)$ performed in within the $R$-matrix
method.

 The resulting ANC (NVC) value  for $\alpha +{\rm {^3He}}\to{\rm
{^7Be}} $(g.s.)  obtained by us is in good agreement with the value
$C_{1\,\,3/2}^2$=25.3 fm$^{-1}$ ($|G_{1\,\,3/2}|^2$=1.20 fm) and
that for $\alpha +{\rm {^3He}}\to{\rm {^7Be}} $(0.429 MeV) differs
noticeably  from the value $C_{1\,\,1/2}^2$=22.0 fm$^{-1}$
($|G_{1\,\,1/2}|^2$=1.04 fm), which were obtained in \cite{Wall84}
within the ($\alpha +^3{\rm {He}}$)-channel resonating-group method.
Also, the results  of the present work  differ noticeably from the
values $C_{1\,\,3/2}^2$=12.6$\pm$ 1.1 fm$^{-1}$ and
$C_{1\,\,1/2}^2=8.41\pm 0.58$ fm$^{-1}$($C_{1\,\,3/2}$=3.55$\pm$
0.15 fm$^{-1/2}$, $C_{1\,\,1/2}$=2.90$\pm$ 0.10 fm$^{-1/2}$,
$|G_{1\,\,3/2}|^2$=0.596$\pm$ 0.052 fm and
$|G_{1\,\,1/2}|^2$=0.397$\pm$ 0.030 fm) \cite{Nol01} as well as
those   $C_{1\,\,3/2}^2=C_{1\,\,1/2}^2$=14.4 fm$^{-1}$
($C_{1\,\,3/2}=C_{1\,\,1/2}$=3.79 fm$^{-1/2}$ and
$|G_{1\,\,3/2}|^2=|G_{1\,\,1/2}|^2$=0.680 fm) \cite{PD}. In this
connection   one would like to note that in \cite{Nol01} the bound
state wave functions, which correspond to   the binding energy for
${\rm {^7Be}}$(g.s.) in the ($\alpha +{\rm {^3He}}$)-channel
differing noticeably from the experimental ones (see Table I in
Ref.\cite{Nol01}), and the initial state wave functions  were
computed with different potentials and, so, these calculations were
not self-consistent.   Since the ANCs   for  $ {\rm {^3He}}+\alpha
\rightarrow {\rm {^7Be}}$ are sensitive to the form of the NN
potential, it is desirable, firstly, to calculate the wave functions
of the bound state using other forms of the NN potential, and,
secondly, in order to  guarantee the self-consistency, the same
forms of the NN potential should be used for such calculation of the
initial wave functions.

\subsection{$\alpha$-particle spectroscopic factors for the mirror
($^7{\rm {Li}}^7{\rm {Be}}$)--pair}

\hspace{0.7cm}The $^{\glqq}$indirectly determined\grqq\, values of
the ANCs for $^3{\rm {He}}+\alpha\rightarrow {\rm {^7Be}}$ presented
in the last line of Table \ref{table2}) and those for     $ \alpha
+t\rightarrow {\rm {^7Li}}$ deduced in Ref.\cite{Igam07} can be used
for obtaining information on the ratio $R_{Z;j_f}=Z_{1j_f}(^7{\rm
{Be}})/Z_{1j_f}({\rm {^7Li}})$ for the virtual $\alpha$ decays of
the bound mirror (${\rm {^7Li}}^7{\rm {Be}}$)-pair, where
$Z_{1j_f}({\rm {^7Be}}) (Z_{1j_f}({\rm {^7Li}}))$ is the
spectroscopic factor for ${\rm {^7Be}}$ (${\rm {^7Li}}$) in the
($\alpha +{\rm {^3He}}$)(($\alpha +t$))-configuration. For this aim,
from $C_{1\,j_f}({\rm {B}})= Z_{1\,j_f}^{1/2}({\rm
{B}})C^{(sp)}_{1\,j_f}({\rm {B}})$ (${\rm {B}}={\rm {^7Li}}$ and
${\rm {^7Be}}$) we form the relation
\begin{equation}
R_{Z;\,j_f}=\frac{R_{C;\,j_f}}{R_{C^{(sp)};\,j_f}},
 \label{18acd}
\end{equation}
where $R_{C;\,j_f}=\Big (C _{1\,j_f}(^7{\rm {Be}})/C_{1\,j_f}({\rm
{^7Li}})\Big)^2$($R_{C^{(sp)};\,j_f}=\Big (C _{1\,j_f}^{(sp)}(^7{\rm
{Be}})/C_{1\,j_f}^{(sp)}({\rm {^7Li}})\Big)^2$) is the ratio of
squares of the ANCs (single-particle ANCs) for the bound mirror
(${\rm {^7Li}}^7{\rm {Be}}$)-pair and  $j_f$=3/2(1/2) for the ground
(first excited) state of the mirror nuclei. Besides, it should be
noted that the relation (\ref{18acd}) allows one to verify a
validity  of the approximation ($R_{C;j_f}\approx
R_{C^{(sp)};\,j_f}$, i.e. $R_{Z;\,j_f}\approx$ 1) used in
Refs.\cite{Tim07} for the mirror (${\rm {^7Li}}^7{\rm {Be}}$)
conjugated $\alpha$ decays.

For the bound  and first excited state of the mirror (${\rm
{^7Li}}^7{\rm {Be}}$)-pair the values of $C _{1\,j_f}^{(sp)}(^7{\rm
{Be}})$ and $C_{1\,j_f}^{(sp)}({\rm {^7Li}})$ change   by the factor
of 1.3 under the variation of  the geometric parameters ($R$ and
$a$) of the adopted Woods--Saxon potential \cite{Neu75, Neu83}
within the aforesaid ranges, while the ratios $R_{C^{(sp)};\,3/2}$
 and $R_{C^{(sp)};\,1/2}$ for the bound  and first excited states of the
mirror (${\rm {^7Li}}^7{\rm {Be}}$)-pair change by only about 1.5\%
and 6\%, respectively. It is seen that the ratios do not depend
practically from variation of the free parameters $R$ and $a$, which
  are equal to
$R_{C^{(sp)};\,3/2} $=1.37$\pm$ 0.02 and
$R_{C^{(sp)};\,1/2}$=1.40$\pm$ 0.09. They are in good agreement with
those calculated in \cite{Tim07} within the microscopic cluster and
two-body potential models (see Table I there). The ratios for the
ANCs are $ R_{C;\,3/2} $=1.83$^{{\rm {+0.18}}}_{{\rm {-0.25}}}\,$
and $R_{C;\,1/2}$=1.77$^{{\rm {+0.19}}}_{{\rm {-0.24}}}\,$. From
(\ref{18acd}) the values of the ratio $R_{Z;\,j_f}$ are equal to
$R_{Z;\,3/2}$=1.34$^{{\rm { +0.13}}}_{{\rm {-0.18}}}\,$ and
$R_{Z;\,1/2}$=1.26$^{{\rm {+0.16}}}_{{\rm {-0.19}}}\,$ for the
ground and the first excited states, respectively. Within their
uncertainties, these values differ slightly from those of
$R_{Z;\,3/2}$=0.995$\pm$0.005 and $R_{Z;\,1/2}$=0.990 calculated in
Ref.\cite{Tim07} within the microscopic cluster model. One notes
that the  values of $R_{Z;\,j_f}$ calculated in \cite{Tim07}  are
sensitive to the model assumptions (the choice of the oscillation
radius $b$ and  the form of the effective $NN$ potential) and such
model dependence may actually influence   the mirror symmetry   for
the $\alpha$-particle spectroscopic factors. The mirror symmetry
breakup for the $\alpha$-particle spectroscopic factors can also be
signalled by the results for the ratio of $S_{34}({\rm
{^7Be}})/S_{34}({\rm {^7Li}})$ at zero energies  for the mirror
(${\rm {^7Li}}^7{\rm {Be}}$)-pair obtained in \cite{Kaj86} within
the resonating-group method by using the seven forms for the
effective $NN$ potential. As shown in \cite{Kaj86}, this ratio is
sensitive to a form of the effective $NN$ potential used and changes
from 1.0   to 1.18 times in a dependence from   the effective $NN$
potential used. One of the possible reasons of   the sensitivity
observed in \cite{Kaj86} can apparently be associated with a
sensitivity of the ratio $R_{Z;\,j_f}$   to a form of the effective
$NN$ potential used. In a contrast of such model dependence observed
in \cite{Kaj86,Tim07}, the problem of the ambiguity connected with
the model $(R,a)$-dependence for the values of the ratios
$R_{Z;\,j_j}$ found by us  from Eq.(\ref{18acd}) is reduced to
minimum within the experimental uncertainty.

It is seen from here   that   the empirical  values of $R_{Z;\,j_f}$
  exceed   unity both for the ground state and for the
first excited state of the mirror (${\rm {^7Li}}^7{\rm {Be}}$)-pair.
This result for  $R_{Z;\,j_f}$ is not accidental  and can be
explained qualitatively by the following consideration. The fact is
that the spectroscopic factor $Z_{1\,j_f}({\rm {^7Li}})$ (or
$Z_{1\,j_f}(^7{\rm {Be}})$)   is determined as a norm of the radial
overlap function of the bound state wave functions of the $t$,
$\alpha$  and ${\rm {^7Li}}$ (or ${\rm {^3He}}$, $\alpha$  and ${\rm
{^7Be}}$) nuclei and is given by Eqs.(100)  and  (101) from Ref.
\cite{Blok77}. The interval of integration ($0\le r<\infty$) in Eq.
(101) can be divided in two parts. In the first integral denoted  by
$Z_{1\,j_f}^{(1)}({\rm {^7Li}})$ for ${\rm {^7Li}}$ and
$Z_{1\,j_f}^{(1)}({\rm {^7Be}})$ for ${\rm {^7Be}}$, the integration
over $r$ covers the region 0$\le r\le r_c$ (the internal region),
where nuclear ($\alpha t$ or $\alpha ^3{\rm {He}}$) interactions are
dominate over the Coulomb interactions. In the second integral
\begin{equation}
Z_{1\,j_f}^{(2)}({\rm {B}})=C _{1\,j_f}^2({\rm
{B}})\int_{r_c}^\infty drW_{\eta_{{\rm B}};\,3/2}^2(2\kappa_{\alpha
a} r),
 \label{18ace}
\end{equation}
 where in the asymptotic region
the radial overlap function entering  the integrand is replaced by
the appropriate Whittaker function (see, for example, Eq.(108) of
Ref.\cite{Blok77}),  interaction between   $a$ and $\alpha$-particle
( $a=t$ for ${\rm B}={\rm {^7Li}}$ or $a={\rm {^3He}}$ for ${\rm
B}={\rm {^7Be}}$) is governed by the Coulomb forces only (the
external region). In (\ref{18ace}), $\kappa_{\alpha
a}=\sqrt{2\mu_{\alpha a}\varepsilon_{\alpha a}}$ and $W_{\eta_{{\rm
{B}}} ;\,3/2}(x)$ is the Whittaker function. One notes that the
magnitudes $Z_{1\,j_f}^{(1)}({\rm {^7Li}})$ ($Z_{1\,j_f}^{(1)}({\rm
{^7Be}})$) and $Z_{1\,j_f}^{(2)}({\rm {^7Li}})$
($Z_{1\,j_f}^{(2)}({\rm {^7Be}})$) define the probability of finding
$t$ (or ${\rm {^3He}}$) in the ($\alpha +t$) configuration (or the
($\alpha +{\rm {^3He}}$) configuration) at distances of $r\le r_c$
and of $r>r_c$, respectively. Obviously $Z_{1\,j_f}({\rm
{^7Li}})=Z_{1\,j_f}^{(1)}({\rm {^7Li}})+Z_{1\,j_f}^{(2)}({\rm
{^7Li}})$ and $Z_{1\,j_f}({\rm {^7Be}})=Z_{1\,j_f}^{(1)}({\rm
{^7Be}})+Z_{1\,j_f}^{(2)}({\rm {^7Be}})$.

An information about values of $Z_{1\,j_f}^{(2)}({\rm {^7Li}})$ and
$Z_{1\,j_f}^{(2)}({\rm {^7Be}})$ can be obtained from (\ref{18ace})
by using the values of the ANCs for $\alpha +t\to{\rm {^7Li}}$ and
$\alpha + {\rm {^3He}}\to{\rm {^7Be}}$ recommended in \cite{Igam07}
and in the present work, respectively. For example, for $r_c\approx
$4.0 fm (the surface regions for the mirror ($^7{\rm {Li}}^7{\rm
{Be}}$)-pair) the calculation shows that the ratio
$R_{Z;\,j_f}^{(2)}=Z_{1\,j_f}^{(2)}({\rm
{^7Be}})/Z_{1\,j_f}^{(2)}({\rm {^7Li}})$ is equal to 1.43$^{{\rm
{+0.13}}}_{{\rm {-0.18}}}$ (1.31$^{{\rm {+0.14}}}_{{\rm {-0.18}}}$)
for the ground (excited) states of the ${\rm {^7Li}}$ and ${\rm
{^7Be}}$ nuclei, i.e. the ratio $R_{Z;\,j_f}^{(2)}>1$. Owing to the
principle of equivalency of nuclear interactions between nucleons of
the ($\alpha t$)-pair in the nucleus ${\rm {^7Li}}$ and ($\alpha
^3{\rm {He}}$)-pair in the nucleus ${\rm {^7Be}}$ \cite{Tim07}, the
values of $Z_{1\,j_f}^{(1)}({\rm {^7Li}})$ and
$Z_{1\,j_f}^{(1)}({\rm {^7Be}})$ should   not differ appreciably. If
one suggests that $R_{Z;\,j_f}^{(1)}\approx$1, then the ratio
$R_{Z;\,j_f}>$1.

\subsection{    The  $^3{\rm {He}}(\alpha,\gamma )^7{\rm {Be}}$
astrophysical $S$ factor at solar energies}

\hspace{0.7cm}  The equation (A1) and the weighted means of the ANCs
obtained  for the $^3{\rm {He}}+\alpha\to{\rm {^7Be}}$(g.s) and
$^3{\rm {He}}+\alpha\to{\rm {^7Be}}$(0.429 MeV) can be used for
extrapolating the $^3{\rm {He}}(\alpha,\gamma )^7{\rm {Be}}$
astrophysical $S$ factor for capture to the ground and first excited
states as well as the total astrophysical $S$ factor at solar
energies ($E\leq 25$ keV). We tested again the fulfilment of the
condition (A2) in the same way as it is done above for E$\ge$ 90 keV
and  similar results plotted in Fig.\ref{fig1} are also observed  at
energies of E$< 90$ keV.

\begin{figure}[!h]
\epsfxsize=14.cm \centerline{\epsfbox{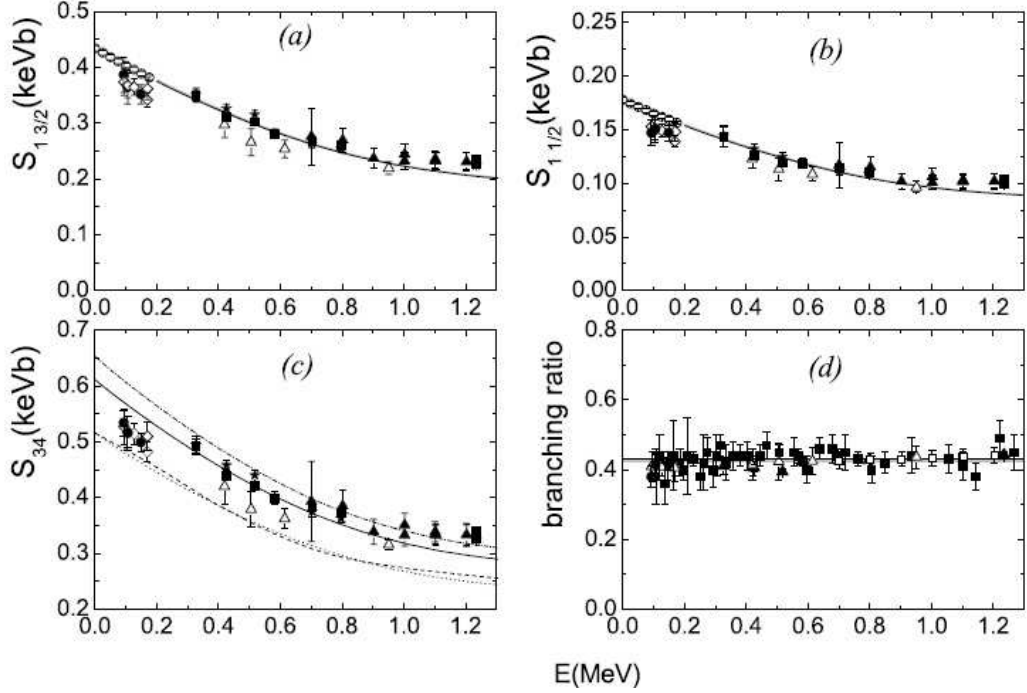}}
\caption{\label{fig6}The astrophysical $S$ factors for the $^3{\rm
{He}}(\alpha,\gamma)^7{\rm {Be}}$(g.s.)  (\emph{a}), $^3{\rm
{He}}(\alpha,\gamma)^7{\rm {Be}}$ (0.429 MeV)   (\emph{b}) and
$^3{\rm {He}}(\alpha,\gamma)^7{\rm {Be}}$ (0.429 MeV)  (\emph{c})
reactions as well as the branching ratio (\emph{d}).  In  (\emph{a})
and (\emph{b}): the open diamond and triangle symbols (the filled
triangle symbols)
 are our result separated
from the total experimental astrophysical $S$ factors of Refs.[7-9]
and \cite{Nara04}, respectively,  (\cite{Di2009}); the filled circle
symbols (filled star and square symbols) are experimental data of
Ref.\cite{Con07} (Ref.\cite{Brown07}, the activation   and the
prompt method, respectively); the open circle symbols are our
results of the extrapolation; in (\emph{c}), the symbols  are data
of all experiments \cite{Nara04}--\cite{Brown07} and the present
work(the open cycle symbols).
  The solid lines present our
calculations performed with the standard values of geometric
parameters $R$=1.80 fm and $a$=0.70 fm.  The dashed line is the
result of \cite{Moh93,Moh2009}. In (\emph{d}):the filled circle,
triangle and square symbols are experimental data taken from
Refs.\cite{Bem06,Con07}, \cite{Brown07} and \cite{Kra82},
respectively, and the open triangle  and square  symbols are our
results. The straight line (width of band ) is our result for the
weighted mean(its uncertainty).}
\end{figure}

 The experimental and calculated astrophysical $S$  factors for
the $^3{\rm {He}}(\alpha,\gamma)^7{\rm {Be}}$(g.s.), $^3{\rm
{He}}(\alpha,\gamma)^7{\rm {Be}}$ (0.429 MeV) and $^3{\rm
{He}}(\alpha,\gamma)^7{\rm {Be}}$ (g.s.+0.429 MeV) reactions are
presented in Table \ref{table1} and displayed in
Fig.\ref{fig6}\emph{a}, \emph{b} and \emph{c}, respectively.   In
Figs.\ref{fig6}\emph{a} and \emph{b}, the open diamond and triangle
symbols (the filled triangle symbols) show our results for the
$^3{\rm {He}}(\alpha,\gamma)^7{\rm {Be}}$(g.s.) and $^3{\rm
{He}}(\alpha,\gamma)^7{\rm {Be}}$(0.429 MeV) reactions (see Table
\ref{table1} also), which are obtained from the analysis of the
total experimental astrophysical $S$ factors of [7--9] and
\cite{Nara04}(\cite{Di2009}), respectively, by using the
corresponding values of the ANCs for each energy $E$ experimental
point presented Figs.\ref{fig5}\emph{a} and \emph{b} ( see Table
\ref{table1} too). There  the open circle symbols lying along the
smooth solid lines are the results of the extrapolation obtained by
us in which each of the quoted uncertainties is the uncertainty
associated with that for the ANCs adopted. All these results are the
second ones of the present work. In Figs.\ref{fig6}\emph{a} and
\emph{b} the experimental data plotted by the filled circle symbols
(filled star and square symbols) are taken from \cite{Con07} (from
\cite{Brown07}).  The solid lines present  our calculations
performed also with the standard values of geometric parameters
$R$=1.80 fm and $a$=0.70 fm.  In Fig.\ref{fig6}\emph{c}, the symbols
are data of all experiments [6--11] and the solid line presents our
calculations performed with the standard values of geometric
parameters $R$=1.80 fm and $a$=0.70 fm by using the weighted means
of  the ANCs ($C_{1\,\,3/2}^2$ and $C_{1\,\,1/2}^2$) presented in
the last line of Table \ref{table2}. There the dashed (dot-dashed)
lines are the results of calculation obtained by using the aforesaid
lower (upper) limit values of the ANCs pointed  out in the last line
of Table \ref{table2} and   the standard values of the geometric
parameters ($R$ and $a$), and the dotted line is the result of
Ref.\cite{Moh93,Moh2009}.   As it is seen from these figures, the
equations (A1), (A4)  and (A5) allow one to perform a correct
extrapolation of the corresponding astrophysical $S$ factors at
solar energies. But, the noticeable systematical underestimation
between the results of calculations performed in
Ref.\cite{Moh93,Moh2009} in respect to the experimental data occurs.

 The weighted means of
 the total astrophysical $S$  factor $S_{34}(E)$
 at solar energies ($E$=0 and 23 ${\rm {keV\,\,b}}$) obtained  by us
  are presented in the the last line of Table \ref{table2}. As it is
 seen from Table \ref{table2} the weighted means of   $S_{34}(0)$,
 deduced  by us   separately from each the activation and the prompt method of
the experimental data   from works  [6--9] and \cite{Brown07,Di2009}
(the first and second lines as well as the fifth and sixth lines),
agree well within their uncertainties with each other and with those
recommended in [9--11]. But, these weighted means of $S_{34}(0)$
obtained by us from the independent analysis of the different data
(the set I and the set II)    differ also noticeably from one
another (about 11\%) and  this distinction  is mainly associated
with  the aforesaid difference observed in magnitudes of the
corresponding ANCs presented in the third and seventh lines of Table
\ref{table2}. Nevertheless, the weighted mean of
$S_{34}$(0)=0.613$^{{\rm {+0.\,026}}}_{{\rm {-0.\,063}}}$ ${\rm
{keV\,\, b}}$, obtained by us by using the weighted means of the
ANCs presented in the last line of Table \ref{table2}, within the
asymmetric uncertainty, which is caused with the asymmetric
uncertainty for the ANCs presented in the last line of Table
\ref{table2}, agrees also with that recommended in [9--11,38]. But,
it is interesting to note that the  central value  of it is closer
to that given in the third line of Table \ref{table2}, than to the
central value of the weighted mean given in the seventh line of
Table \ref{table2}.   Also, the  astrophysical $S$ factors,
calculated by using the values of the ANCs obtained  separately from
the set I, the set II and both them (see Table \ref{table2}), are
fitted independently  using a second-order polynomial within three
energy intervals (0$\le E\le$ 500 keV , 0$\le E\le$ 1000 keV and
0$\le E\le$ 1200 keV). The results for the slop
$S_{34}^{\prime}$(0)/$S_{34}$(0) are to be -0.711 MeV$^{-1}$, -0.734
MeV$^{-1}$ and  -0.726 MeV$^{-1}$ in dependence on the aforesaid
intervals, respectively, and they do not depend on the values of the
ANCs used. One notes that they also are in agreement with -0.73
MeV$^{-1}$ \cite{Moh2009} and -0.92$\pm$0.18 MeV$^{-1}$
\cite{Cyb2008}.  It is seen  from here that   the $S_{34}(E)$
calculated  by us (the solid lines in Fig.\ref{fig6}\emph{a},
\emph{b} and \emph{c}) and that obtained in \cite{Moh2009,Cyb2008}
have  practically the same energy dependence within the aforesaid
energy interval  but they differ mainly with each other by  a
normalization.

Our result for $S_{34}(0)$ differs   noticeably on  that recommended
in Refs.\cite{Nol01} \cite{Igam97} and \cite{PD}($\approx$ 40 keVb,
0.52$\pm$0.03 keVb and  0.51$\pm$0.04 keV b, respectively). This
circumstance is apparently connected with  the underestimation of
the contribution from the external part in the amplitude admitted in
these works. Besides, the result of the present work is noticeably
larger than the result of $S_{34}(0)$=0.516 (0.53) keV b
\cite{Moh93}(\cite{Moh2009}) obtained within the standard two-body
($\alpha +{\rm {^3He}}$) potential by using $\alpha{\rm {^3He}}$
potential deduced by a   double-folding procedure. One of the
possible reason of this discrepancy can be apparently associated
with the assumption admitted in \cite{Moh93,Moh2009} that  a value
of the ratio $R_{Z;\,j_f}$ for the bound   mirror (${\rm {^7Li}}{\rm
{^7Be}}$)-pair is taken equal to unity ($Z_{1\,3/2}({\rm {^7Be}})$ =
$Z_{1\,1/2}({\rm {^7Be}})$=1.17 \cite{Kurath75}), which in turn
 results in the observed systematical  underestimation of the
calculated $S_{34}(E)$ in respect to the experimental $S$ factors
(see the dashed line in Fig.\ref{fig6}\emph{c}). But, as shown in
Subsection 2.3 the values of $R_{Z;\,j_f}$ for the ground and first
excited states of the mirror (${\rm {^7Li}}{\rm {^7Be}}$)-pair are
  large unity. Therefore, the underestimated values of
$Z_{1\,3/2}({\rm {^7Be}})$ and $Z_{1\,1/2}({\rm {^7Be}})$  used in
\cite{Moh93} also result in the underestimated value of $S_{34}(0)$
for the direct capture ${\rm {^3He}}(\alpha,\gamma ){\rm {^7Be}}$
reaction. Perhaps, the assumption about equal values of the
spectroscopic factor ($Z_{1\,3/2}({\rm {^7Li}})$ = $Z_{1\,1/2}({\rm
{^7Li}})$=1.17 \cite{Moh93,Kurath75}) is correct only for the
spectroscopic factors $Z_{1\,j}({\rm {^7Li}})$ since the value of
$S_{34}(0)$ obtained in \cite{Igam07} and \cite{Moh93}
 for the direct capture $t(\alpha,\gamma ){\rm {^7Li}}$ reaction   agree excellently
 with each other. One notes that in Ref.\cite{Igam07} the analysis
 of
the   $t(\alpha,\gamma ){\rm {^7Li}}$ experimental astrophysical $S$
factors \cite{Brune1999}  has also been performed within MTBPA and
the ANC for $\alpha +t\to{\rm {^7Li}}$(g.s.), which has been
 deduced there from the results of Ref.\cite{Moh93}, also is in good agreement with that recommended
by authors of  Ref.\cite{Igam07}.

Nevertheless, we observe that the value of $S_{34}(0)$= 0.56 ${\rm
{keV\,\, b}}$ \cite{Lan86}   obtained within the microscopical
($\alpha +{\rm {^3He}}$)-cluster approach    is   in    agreement
with our result within the uncertainty. Besides, our result is also
in excellent agreement with that of $S_{34}(0)$= 0.609 ${\rm
{keV\,\, b}}$ \cite{Kaj86} and $S_{34}(0)$= 0.621 ${\rm {keV\,\,
b}}$ \cite{Wall84} obtained within the   ($\alpha +{\rm
{^3He}}$)-channel of version of  the resonating-group method by
using the modified Wildermuth-Tang (MWT) and  the near-Serber
exchange mixture   forms for the effective NN potential,
respectively. It follows from here that the mutual agreement between
the results obtained in the present work and works of
\cite{Kaj86,Lan86,Wall84}, which is based on the common
approximation about the cluster $(\alpha+{\rm {^3He}})$ structure of
the ${\rm {^7Be}}$, allows one to draw a conclusion about  the
dominant contribution of the $(\alpha+{\rm {^3He}})$ clusterization
to the low-energy ${\rm {^3He}}(\alpha,\gamma){\rm {^7Be}}$ cross
section both in the absolute normalization  and in the energy
dependence [6--11]. Therefore, single-channel $(\alpha+{\rm
{^3He}})$ approximation for ${\rm {^7Be}}$ \cite{Kaj86,Lan86} is
quite appropriate for this reaction in the considered energy range.

Also, it is interesting to note  that the ratios of  the
$^{\glqq}$indirectly determined\grqq\, astrophysical $S$  factors,
$S_{1\,\,3/2}$(0) and $S_{1\,\,1/2}(0)$, for the $^3{\rm
{He}}(\alpha,\gamma )^7{\rm {Be}}$ reaction populating to the ground
and first excited states obtained in the present work to those for
the mirror $t(\alpha,\gamma )^7{\rm {Li}}$ reaction populating to
the ground and first excited states deduced in Ref.\cite{Igam07} are
equal to $R_S^{(g.s.)}$=6.87$^{+0.70}_{-0.87}$ and
$R_S^{(exc)}$=6.11$^{+0.67}_{-0.86}$ , respectively. These values
are in a good agreement with those of $R_S^{(g.s.)}$=6.6 and
$R_S^{(exc)}$=5.9 deduced in Ref.\cite{Tim07} within the microscopic
cluster model. This result also confirms directly our estimation for
the ratio $R_{C;\,j_f}$ obtained above since the ANCs for
$t+\alpha\to{\rm {^7Li}}$(g.s)  and $t+\alpha\to{\rm {^7Li}}$(0.478
MeV) as well as  the ANCs for ${\rm {^3He}}+\alpha\to{\rm
{^7Be}}$(g.s)  and ${\rm {^3He}}+\alpha\to{\rm {^7Be}}$(0.429 MeV)
determine the astrophysical $S$ factors for the $t(\alpha,\gamma
)^7{\rm {Li}}$ and $^3{\rm {He}}(\alpha,\gamma )^7{\rm {Be}}$
reactions at zero energies and, consequently, the ratios
$R_S^{(g.s.)}$ and $R_S^{(exc)}$ are proportional to $R_{C;\,3/2}$
and $R_{C;\,1/2}$, respectively.

Fig. \ref{fig6}\emph{d}  shows a comparison  between the branching
ratio $R^{exp}(E)$ obtained in the present work (the open triangle
and square  symbols) and that recommended in Refs.\cite{Kra82} (the
filled square symbols), in \cite{Bem06,Con07} (the filled circle
symbols) and
  \cite{Brown07} (the filled triangle symbols). The weighted mean
$\bar{R}^{exp}$ of the $R^{exp}(E)$    recommended by us  is equal
to   $\bar{R}^{exp}$=0.43$\pm$ 0.01. As it is seen from
Fig.\ref{fig6}\emph{d}, the branching ratio obtained in the present
work and in \cite{Bem06,Con07,Brown07} is in a good agreement with
that recommended in Ref.\cite{Kra82} although the underestimation
occurs for the $S_{34}^{exp}(E)$ obtained in Ref.\cite{Kra82}. Such
a good agreement between two of the experimental data for the
$R^{exp}(E)$ can apparently be explained by the fact that  there is
a reduction factor  in \cite{Kra82}, being overall for the $^3{\rm
{He}}(\alpha,\gamma){\rm {^7Be}}$(g.s.) and $^3{\rm
{He}}(\alpha,\gamma){\rm {^7Be}}$(0.429 MeV) astrophysical $S$
factors. The present result for $\bar{R}^{exp}$  is in an excellent
agreement with those of 0.43$\pm$ 0.02 \cite{Kra82} and 0.43
\cite{Moh93,Alt88} but  is noticeably larger than 0.37 \cite{Nol01}
and 0.32$\pm$ 0.01 \cite{Sch87}.

\section{Conclusion}

\hspace{0.7cm} The analysis of the modern experimental astrophysical
$S$ factors, $S^{exp}_{34}(E)$, for the $^3{\rm
{He}}(\alpha,\gamma)^7{\rm {Be}}$ reaction, which were precisely
measured at energies $E$=92.9-1235 keV [6--11], has been performed
within the modified two-body potential approach \cite{Igam07}. The
performed scrupulous  analysis shows quantitatively  that the
$^3{\rm {He}}(\alpha,\gamma)^7{\rm {Be}}$ reaction within the
considered energy ranges is mainly peripheral and the
parametrization of the direct astrophysical $S$ factors in terms of
ANCs for the $^3{\rm {He}}+\alpha\rightarrow {\rm {^7Be}}$ is
adequate to the physics of the peripheral reaction under
consideration.

It is shown that the experimental astrophysical $S$  factors of the
reaction  under consideration  [6--11]  can be used as an
independent source of getting the information about the ANCs (or
NVCs) for $^3{\rm {He}}+\alpha\rightarrow\rm {^7Be}$ (or for the
virtual decay $\rm {^7Be}\rightarrow\alpha +^3{\rm {He}}$), and  the
found ANCs can predict the  experimental astrophysical $S$ factors
separated for the ${\rm {^3He}}(\alpha,\gamma ){\rm {^7Be}}$(g.s.)
and ${\rm {^3He}}(\alpha,\gamma ){\rm {^7Be}}$(0.429 MeV)  reactions
at low experimentally acceptable energy regions (126.5$\le$ E$\le$
1203 keV) obtained  from  the total experimental astrophysical $S$
factors [6--9,11]. The new estimation for the weighted means of the
ANCs for $^3{\rm {He}}+\alpha\rightarrow{\rm {^7Be}}$ and NVCs for
the virtual decay $\rm {^7Be}\rightarrow\alpha +^3{\rm {He}}$ are
obtained.   Also, the values of ANCs were used for getting the
information about the $\alpha$-particle spectroscopic factors for
the mirror ($^7Li{\rm {^7Be}}$)-pair.

The obtained values  of the ANCs  were also used     for an
extrapolation of astrophysical $S$ factors at energies less than 90
keV, including $E$=0. In particular,  the weighted mean of the
branching ratio $\bar{R}^{exp}$ ($\bar{R}^{exp}$=0.43$\pm$ 0.01) and
the total astrophysical $S$ factor $S_{34}(0)$
($S_{34}$(0)=0.613$^{{\rm {+0.026}}}_{{\rm {-0.063}}}$ ${\rm
{keV\,\, b}}$)  obtained here are in   agreement with that deduced
in [7--11] from the analysis the same experimental asprophysical $S$
factors. Besides, our result for $S_{34}(0)$ is in an agreement with
that $S_{34}(0)$=0.56 keV \cite{Lan86} obtained within the
microscopical single-channel ($\alpha +{\rm {^3He}}$) cluster model,
$S_{34}(0)$= 0.609 ${\rm {keV\,\, b}}$ \cite{Kaj86} and $S_{34}(0)$=
0.621 ${\rm {keV\,\, b}}$ \cite{Wall84} obtained within the ($\alpha
+{\rm {^3He}}$)-channel of version of  the resonating-group method,
but it is noticeably larger than the result of $S_{34}(0)$=0.516
(0.53) keV \cite{Moh93}(\cite{Moh2009}) obtained within the standard
two-body ($\alpha +{\rm {^3He}}$) potential by using $\alpha{\rm
{^3He}}$ potential deduced by a double-folding procedure.

\vspace{5mm} {\bf Acknowledgments} \vspace{5mm}

The authors are deeply grateful to S. V. Artemov,  L.D. Blokhintsev
and A.M. Mukhamedzhanov for discussions and general encouragement.
The authors  thank also  D.Bemmerer for providing the experimental
results of the updated data analysis. The work has been supported by
The Academy of Science  of The Republic of Uzbekistan (Grant
No.FA-F2-F077).

\vspace{5mm} {\bf Appendix: Basic formulae} \vspace{5mm}

\hspace{0.7cm}  Here we repeat only   the idea and the essential
formulae of the MTBPA \cite{Igam07} specialized for the $^3{\rm
{He}}(\alpha,\gamma){\rm {^7Be}}$ astrophysical $S$ factor that are
important for the following analysis.

 According to \cite{Igam07}, for fixed $l_f$ and $j_f$ we can write the astrophysical
 $S$ factor, $S_{l_fj_f}(E)$, for the peripheral
direct capture $^3{\rm {He}}(\alpha,\gamma)^7{\rm {Be}}$ reaction in
the following form
$$
 S_{l_fj_f}(E)=C^2_ {l_fj_f}{\cal
{R}}_{l_fj_f}(E,C_{l_fj_f}^{(sp)}). \eqno(A1)
$$
Here,   $C_ {l_fj_f}$ is the ANC  for  ${\rm {^3He}}+\alpha\to{\rm
{^7Be}} $, which determines the amplitude of the tail of the ${\rm
{^7Be}}$ nucleus bound   state wave function in the ($\alpha+{\rm
{^3He}}$)-channel and is related to the spectroscopic factor
$Z_{l_fj_f}$ for the ($\alpha +{\rm {^3He}}$)-configuration with the
quantum numbers $l_f$ and $j_f$ in the ${\rm {^7Be}}$ nucleus by the
equation $C_{l_fj_f}= Z_{l_fj_f}^{1/2}C^{(sp)}_{l_fj_f}$
\,\cite{Blok77}, and ${\cal {R}}_{l_fj_f}(E,C^{(sp)}_{l_fj_f})=
S_{l_fj_f}^{(sp)}(E)/(C^{(sp)}_{l_fj_f})^2$, where $
S_{l_fj_f}^{(sp)}(E)$  is the single-particle astrophysical $S$
factor \cite{An99} and $C^{(sp)}_{l_fj_f}$ ($\equiv
C^{(sp)}_{l_fj_f}(R,a)$ \cite{Gon82}) is the single-particle ANC,
which determines the amplitude of the tail of the single-particle
wave function of the bound ($\alpha+^3{\rm {He}}$) state
$\varphi_{l_fj_f}(r)$($\equiv \varphi_{l_fj_f}(r;
C^{(sp)}_{l_fj_f}$) \cite{Gon82}) and in turn is itself a function
of the geometric parameters (radius  of $R$ and diffuseness $a$) of
the Woods-Saxon potential, i.e.  $C^{(sp)}_{l_fj_f}\equiv
C^{(sp)}_{l_fj_f}(R,a)$\cite{Gon82}.

In order to  make the   dependence of the ${\cal
{R}}_{l_fj_f}(E,C_{l_fj_f}^{(sp)})$ function on $C_{l_fj_f}^{(sp)}$
more explicit,   in the radial matrix element \cite{Igam07,Robert81}
entering in  the ${\cal {R}}_{l_fj_f}(E,C_{l_fj_f}^{(sp)})$
function, we  split  the space of interaction   in  two parts
separated by the channel radius $r_c$: the interior part ($0\le r\le
r_c)$, where nuclear forces between the $\alpha{\rm{^3He}}$-pair are
important, and the exterior part ($r_c\le r<\infty$), where the
interaction between the $\alpha$-particle and ${\rm{^3He}}$ is
governed by Coulomb force only.  The contribution from the exterior
part of the radial matrix element into the ${\cal
{R}}_{l_fj_f}(E,C_{l_fj_f}^{(sp)})$ function does not depend on
$C_{l_fj_f}^{(sp)}$ since for $r>r_c$ the wave function
 $\varphi_{l_fj_f}(r;C_{l_fj_f}^{(sp)})$   can be
approximated by its asymptotic behavior \cite{Blok77}. Consequently,
the parametrization of the astrophysical $S$ factor in the form (A1)
makes one it possible to fix a contribution from the exterior region
($r_c\le r<\infty$), which is dominant for the peripheral reaction,
by a model independent way if the ANCs $C_{l_fj_f}^2$ are known. It
follows from here that the contribution from the interior part of
the radial matrix element into the ${\cal
{R}}_{l_fj_f}(E,C_{l_fj_f}^{(sp)})$ function, which depends on
$C_{l_fj_f}^{(sp)}$ through   the fraction
$\varphi_{l_fj_f}(r;C_{l_fj_f}^{(sp)})/C_{l_fj_f}^{(sp)}$
 \cite{Gon82,Mukh2005}, must exactly determine the
dependence of the ${\cal {R}}_{l_fj_f}(E,C_{l_fj_f}^{(sp)})$
function  on $C_{l_fj_f}^{(sp)}$.

 In Eq. (A1)  the ANCs $C_{l_fj_f}^2$ and the free parameter $C_{l_fj_f}^{(sp)}$
are   unknown. But, for the peripheral $^3{\rm
{He}}(\alpha,\gamma){\rm {^7Be}}$ reaction the equation (A1) can be
used for determination of the ANCs. For this aim, obviously the
following additional requirements \cite{Igam07}
 $$ {\cal
{R}}_{l_fj_f}(E,C^{(sp)}_{l_fj_f})=f(E) \eqno(A2)
$$
and
$$
C_{l_fj_f}^2=\frac{S_{l_fj_f}(E)}{{\cal{R}}_{l_fj_f}(E,C_{l_fj_f}^{(sp)})}
=const \eqno(A3)
$$
must be  fulfilled as a function of the free parameter
$C_{l_fj_f}^{(sp)}$ for each energy  $E$ experimental point from the
range $E_{min}\le E\le E_{max}$ and values of the function of ${\cal
{R}}_{l_fj_f}(E,C_{l_fj_f}^{(sp)})$ from (A2).

The fulfillment of     the relations (A2) and (A3) or their
violation within the experimental uncertainty for
$S_{l_fj_f}^{exp}(E)$ enables one, firstly, to determine  an
interval for energies E where the dominance of extra-nuclear capture
occurs and, secondly,  to obtain the value $(C_{l_fj_f}^{exp})^2$
for $^3{\rm {He}}+\alpha\to{\rm {^7Be}} $ using the  experimental
astrophysical $S$ factors $S^{exp}_{l_fj_f}(E)$, precisely measured
by authors  of Refs. [6--11], instead of $S_{l_fj_f}(E)$, i.e.
$$
(C_{{l_fj_f}}^{exp})^2=\frac{S^{exp}_{l_fj_f}(E)} {{\cal
{R}}_{l_fj_f}(E,C_{l_fj_f}^{(sp)})}. \eqno(A4)
$$
  Then, the value   $(C_{l_fj_f}^{exp})^2$ can be used for
extrapolation of
 the astrophysical $S$ factor $S_{l_fj_f}(E)$ to the region of
 experimental inaccessible energies
0$\le E<E_{min}$ by using the obtained value $(C_{l_fj_f}^{exp})^2$
in (A1).

   The total astrophysical $S$  factor for the $^3{\rm
{He}}(\alpha,\gamma)^7{\rm {Be}}$(g.s.+0.429 MeV) reaction is given
by
$$
S_{34}(E)=\sum_{l_f=1;\,j_f=1/2,\,3/2} S_{l_fj_f}(E)=
  \eqno(A5)
$$
$$
= C_{1\,\,3/2}^2[{{\cal R}}_{1\,\,3/2}(E,C_{1\,\,3/2}^{(sp)})+
\lambda_C{{\cal R}}_{1\,\,1/2}(E,C_{1\,\,1/2}^{(sp)})]\eqno(A6)
$$
$$
=C_{1\,\,3/2}^2{{\cal R}}_{1\,\,3/2}(E,C_{1\,\,3/2}^{(sp)})[1+R(E)]
\eqno(A7)
$$
in which $\lambda_C=(C_{1\,\,1/2}/C_{1\,\,3/2})^2$   and  $R(E)$  is
a branching ratio.

One notes that in the two-body potential model the ANC $C_{l_fj_f}$
is related to the NVC $G_{l_fj_f}$ for the virtual decay ${\rm
{^7Be}}\to\alpha + {\rm {^3He}}$  by the equation \cite{Blok77}
$$
G_{l_fj_f}=-i^{l_f+\eta_{{\rm
{\,^7Be}}}}\frac{\sqrt{\pi}}{\mu}C_{l_fj_f}, \eqno(A8)
$$
where $\eta_{{\rm {\,^7Be}}}$ is the Coulomb parameter for the ${\rm
{^7Be}}\,(\alpha+{\rm {^3He}})$ bound state.   In (A8) the
combinatorial factor taking into account the nucleon's identity is
absorbed in $C_{l_fj_f}$ and its   numerical value depends on a
specific model used to describe wave functions of the ${\rm
{^3He}}$, $\alpha$ and ${\rm {^7Be}}$ nuclei \cite{Blokh2008}.
Hence, the proportionality factor in (A8), which relates NVC's and
ANCs, depends on the choice of nuclear model \cite{Blokh2008}. But,
as it is noted in \cite{Blokh2008}, the NVC $G_{l_fj_f}$ is a more
fundamental quantity than the ANC $C_{l_fj_f}$ since the NVC is
determined in a model-independent way as the residue of the partial
$S$-matrix of the elastic $\alpha {\rm {^3He}}$-scattering at the
pole $E= -\varepsilon_{\alpha{\rm {^3He}}}$
($\varepsilon_{\alpha{\rm {^3He}}}$ is the binding energy of the
bound ($\alpha +{\rm {^3He}}$) state of ${\rm {^7Be}}$)
\cite{Blok77,Blokh2008}. Therefore,   it is also of interest  to get
an information about values of the  NVCs from Eqs. (A4) and (A8).

\newpage

\newpage

\newpage

\newpage
\newpage
\newpage
\begin{landscape}
\begin{table}
\caption{\label{table1}The $^{\glqq}$indirectly determined\grqq\,
values of the asymptotic normalization constants
($(C_{13/2}^{exp})^2$ and ($C_{11/2}^{exp})^2$) for
$\,\,\,\,\,\,\,\,\,\,\,\,\,\,\,\,\,\,^3{\rm {He}}+\alpha\to^7{\rm
{Be}}$, the experimental astrophysical $S$  factors
($S^{exp}_{1j_f}$ and $S^{exp}_{34}(E)$) and branching ratio
($R^{exp}(E)$) at different energies $E$.} \vspace{0.2cm}
 \begin{tabular}{|c|c|c|c|c|c|c|}\hline
$E$&\multicolumn{2}{c|}{$(C_{1\,\,j_f}^{exp})^2$}&\multicolumn{2}{c|}{$S^{exp}_{1\,\,j_f}$}&$S^{exp}_{34}(E)$&$R^{exp}(E)$ \\
 (keV)&\multicolumn{2}{c|}{(fm$^{-1})$}&\multicolumn{2}{c|}{(${\rm  {keV\,\,\,b}}$)}&(${\rm  {keV\,\,\,b}}$)& \\
 \cline{2-5}
&$j_f$=3/2&$j_f$=1/2&$j_f$=3/2&$j_f$=1/2&&\\ \hline
92.9$^{*)}$&22.0$\pm$1.8&14.0$\pm$1.2&0.387$\pm$0.031\cite{Bem06,Con07}&0.147$\pm$0.012\cite{Bem06,Con07}&0.534$\pm$0.023\cite{Bem06,Con07}&0.380$\pm$0.030\cite{Bem06,Con07}\\\hline
93.3$^{**)}$&21.4$\pm$1.4&14.7$\pm$0.9&0.374$\pm$0.020&0.153$\pm$0.009&0.527$\pm$0.027\cite{Con07}&0.409$\pm$0.03\\\hline
105.6$^{*)}$&21.0$\pm$1.8&14.6$\pm$1.2&0.365$\pm$0.030\cite{Bem06,Gy07}&0.151$\pm$0.012\cite{Bem06,Gy07}&0.516$\pm$0.031\cite{Bem06,Gy07}&0.415$\pm$0.029\cite{Bem06,Con07}\\\hline
106.1$^{**)}$&21.2$\pm$1.3&14.6$\pm$0.9&0.368$\pm$0.020&0.150$\pm$0.009&0.518$\pm$0.024\cite{Con07}&0.408$\pm$0.020\\\hline
126.5$^{*)}$&21.2$\pm$0.9&14.6$\pm$0.6&0.366$\pm$0.023&0.148$\pm$0.009&0.514$\pm$0.019\cite{Bem06,Gy07}&0.404$\pm$0.020\\\hline
147.7$^{*)}$&20.8$\pm$1.1&14.6$\pm$0.7&0.352$\pm$0.017\cite{Bem06}&0.147$\pm$0.007\cite{Bem06}&0.499$\pm$0.017\cite{Bem06,Gy07}&0.417$\pm$0.020\cite{Bem06}\\\hline
168.9$^{*)}$&20.6$\pm$0.8&14.1$\pm$0.6&0.343$\pm$0.010&0.139$\pm$0.006&0.482$\pm$0.017\cite{Bem06,Gy07}&0.405$\pm$0.020\\\hline
170.1$^{**)}$&21.9$\pm$1.2&15.0$\pm$0.8&0.362$\pm$0.020&0.148$\pm$0.008&0.510$\pm$0.021\cite{Con07}&0.409$\pm$0.030\\\hline
420.0$^{*)}$&21.4$\pm$1.7&14.7$\pm$1.1&0.297$\pm$0.020&0.123$\pm$0.009&0.420$\pm$0.030\cite{Nara04}&0.414$\pm$0.050\\\hline
506.0$^{*)}$&20.9$\pm$1.9&14.3$\pm$1.3&0.266$\pm$0.020&0.113$\pm$0.010&0.379$\pm$0.030\cite{Nara04}&0.424$\pm$0.050\\\hline
615.0$^{*)}$&21.5$\pm$1.4&14.7$\pm$0.9&0.254$\pm$0.020&0.108$\pm$0.006&0.362$\pm$0.020\cite{Nara04}&0.425$\pm$0.040\\\hline
951.0$^{*)}$&22.7$\pm$1.2&15.6$\pm$0.8&0.220$\pm$0.010&0.096$\pm$0.005&0.316$\pm$0.010\cite{Nara04}&0.436$\pm$0.030\\\hline
\end{tabular}

$^{*)}$  the activation\\
$^{**)}$ the prompt method\\
\end{table}
\end{landscape}
\newpage
\begin{landscape}
\begin{table}
\caption{\label{table2}The weighted means of the ANC-values
($C^{{\rm exp}})^2$ for   $^3He +\alpha\rightarrow ^7{\rm Be}$,
NVC's $\mid G\mid^2_{{\rm exp}}$ and the calculated values of $
S_{3\,4}(E)$ at energies $E$=0 and 23 keV. The the second and third
lines (the fifth, sixth  and ninth lines) correspond to the results
obtained by means of the analysis of data from works pointed out in
the first column, and the penultimate   line corresponds to that
obtained by using data from all experiments [6--11]. The figures
parenthetical are the weighted means obtained from corresponding
ones given for the activation and the prompt method.}\vspace{0.2cm}
\begin{tabular}{|l|c|c|c|c|c|c|}\hline
Experimental data&$(C^{{\rm exp}}_{1\,3/2})^2$,&$\mid
G_{1\,3/2}\mid^2_{{\rm exp}}$,&$(C^{{\rm exp}}_{1\,1/2})^2$, &$\mid
G_{1\,1/2}\mid^2_{{\rm exp}}$,&$S_{3\,4}$(0), &$S_{3\,4}$({\rm {23\,\,keV}}), \\
& fm$^{-1}$& fm&fm$^{-1}$& fm&keV
b&keV b\\
 \hline [6--9](the
activation)&21.2$\pm$0.4&1.00$\pm$0.02&14.5$\pm$0.3&0.688$\pm$0.013&0.560$\pm$0.003&0.551$\pm$0.003
\\
\cite{Con07}(the prompt
&21.5$\pm$0.7&1.02$\pm$0.04&14.8$\pm$0.5&0.697$\pm$0.024&0.568$\pm$0.002&0.558$\pm$0.002
\\
method),\, the set I&(21.3$\pm$0.4)&(1.01$\pm$0.02)&(14.6$\pm$0.2)&(0.690$\pm$0.011)&(0.566$\pm$ 0.004)&(0.556$\pm$0.003)\\
&&&&&0.560$\pm$0.017 \cite{Con07}&\\
\hline \cite{Brown07}(the activation)&24.0$\pm $0.4&
1.13$\pm$0.02&16.2$\pm$0.2&0.768$\pm$0.011&0.630$\pm$0.008&0.619$\pm$0.008
\\
\cite{Brown07}(the prompt me-
&24.1$\pm$0.3&1.14$\pm$0.02&16.4$\pm$0.2&0.773$\pm$0.011&0.624$\pm$0.010&$0.612\pm$0.010
\\
thod) and \cite{Di2009} (the
&(24.1$\pm$0.2)&(1.14$\pm$0.01)&(16.3$\pm$0.2)&(0.771$\pm$0.008)&
(0.628$\pm$ 0.006)&(0.616$\pm$0.006)\\
${\rm {^7Be}}$ recoils),the set II&&&&&0.596$\pm$ 0.021 \cite{Brown07}&\\
&&&&&0.57$\pm$ 0.04 \cite{Di2009}&\\
${\rm {[6-10]}}$
&&&&&0.580$\pm$ 0.043 \cite{Cyb2008}&\\
\hline [6--11] &23.3$^{{\rm {+1.0}}}_{{\rm {-2.4}}}$
 &1.10$^{{\rm {+0.05}}}_{{\rm {-0.11}}}$&15.9$^{{\rm
{+0.6}}}_{{\rm {-1.5}}}$&0.751$^{{\rm {+0.028}}}_{{\rm
{-0.072}}}$&0.613$^{{\rm {+0.026}}}_{{\rm {-0.063}}}$&0.601$^{{\rm
{+0.030}}}_{{\rm {-0.072}}}$\\
\hline
\end{tabular}
\end{table}
\end{landscape}

\newpage
\begin{thebibliography} {*}
\bibitem{Bah05}
J. N. Bahcall, A. M. Serenelli, and S. Basu,  Astrophys. J. {\bf
621}, L85 (2005).
\bibitem{Bah82}
J. N. Bahcall, W.F. Huebner, S. H. Lubow, P. D. Parker, and R. K.
Ulrich, Rev.Mod.Phys.  {\bf 54}, 767 (1982).
\bibitem{Bah92}
J. N. Bahcall, and M. H. Pinsonneault, Rev.Mod.Phys. {\bf 64}, 781
(1992).
\bibitem{Adel98}
 E.G. Adelberger, S.M.Austin, J.N. Bahcall, A.B.Balantekin,
 G.Bogaert, L.S.Brown, L.Buchmann, F.E.Cecil, A.E.Champagne, L. de Braeckeleer,
Ch.A. Duba, S.R. Elliott, S.J. Freedman, M. Gai, G.Goldring, Ch. R.
Gould, A. Gruzinov, W.C. Haxton, K.M. Heeger, and E.Henley,  Rev.
Mod. Phys. {\bf 70}, 1265 (1998).
\bibitem{An99}
C.Angulo, M.Arnould, M.Rayet, P.Descouvemont, D.Baye,
C.Leclercq-Willain, A.Coc, S.Barhoumi, P.Aguer, C.Rolfs, R.Kunz,
J.W. Hammer, A.Mayer, T.  Paradellis, S.Kossionides, C.Chronidou,
K.Spyrou, S.Degl'Innocenti, G.  Fiorentini, B.Ricci, S.Zavatarelli,
C.Providencia, H.Wolters, J.Soares, C.Grama, J.Rahighi, A.Shotter,
M.L.Rachti, Nucl. Phys. A {\bf 656}, 3 (1999).
\bibitem{Nara04}
B. S. Nara Singh, M. Hass, Y. Nir-El, and G. Haquin, Phys.Rev.Lett.
{\bf 93}, 262503 (2004).
\bibitem{Bem06}
D. Bemmerer, F. Confortola, H. Costantini, A. Formicola, Gy.
Gy\"{u}rky, R. Bonetti, C. Broggini, P. Corvisiero, Z. Elekes, Zs.
F\"{u}l\"{o}p, G. Gervino, A. Guglielmetti, C. Gustavino, G.
Imbriani, M. Junker, M. Laubenstein, A. Lemut, B. Limata, V.Lozza,
M. Marta, R. Menegazzo, P. Prati, V. Roca, C. Rolfs, C. Rossi
Alvarez, E. Somorjai, O. Straniero, F. Strieder, F. Terrasi, and
H.P. Trautvetter (The LUNA Collaboration), Phys.Rev.Lett. {\bf 97},
122502 (2006); private communication.
\bibitem{Gy07}
Gy. Gy\"{u}rky, F. Confortola, H. Costantini, A. Formicola, D.
Bemmerer, R. Bonetti, C. Broggini, P. Corvisiero, Z. Elekes, Zs.
F\"{u}l\"{o}p, G. Gervino, A. Guglielmetti, C. Gustavino, G.
Imbriani, M. Junker, M. Laubenstein, A. Lemut, B. Limata, V.Lozza,
M. Marta, R. Menegazzo, P. Prati, V. Roca, C. Rolfs, C. Rossi
Alvarez, E. Somorjai, O. Straniero, F. Strieder, F. Terrasi, and
H.P. Trautvetter (The LUNA Collaboration), Phys.Rev. C {\bf 75},
035805 (2007) .
\bibitem{Con07}
F. Confortola, D. Bemmerer,  H. Costantini, A. Formicola, Gy.
Gy\"{u}rky, P. Bezzon, R. Bonetti, C. Broggini, P. Corvisiero, Z.
Elekes, Zs. F\"{u}l\"{o}p, G. Gervino, A. Guglielmetti, C.
Gustavino, G. Imbriani, M. Junker, M. Laubenstein, A. Lemut, B.
Limata, V.Lozza, M. Marta, R. Menegazzo, P. Prati, V. Roca, C.
Rolfs, C. Rossi Alvarez, E. Somorjai, O. Straniero, F. Strieder, F.
Terrasi, and H.P. Trautvetter (The LUNA Collaboration), Phys.Rev. C
{\bf 75}, 065803 (2007).
\bibitem{Brown07}
T.A.D. Brown, C. Bordeanu, R.F. Snover, D.W. Storm, D. Melconian,
A.L. Sallaska, S.K.L. Sjue, S. Triambak, Phys.Rev. C {\bf 76},
055801 (2007).
\bibitem{Di2009} A Di Leva, L. Gialanella, R. Kunz,
D. Rogalla,
 D. Sch\"{u}rmann,
 F. Strieder, M. De Cesare, N. De
Cesare, A. D'Onofrio, Z. F\"{u}l\"{o}p, G. Gy\"{u}rky, G. Imbariani,
G. Mangano, A. Ordine, V. Roca, C. Rolfs, M. Romano, E. Somorjai,
and F. Terrasi, Phys.Rev.Lett. {\bf 102}, 232502 (2009).
\bibitem{Kaj86}
T. Kajino, Nucl.Phys. A {\bf 460}, 559 (1986).
\bibitem{PD}
P.Descouvement, A. Adahchour, C. Angulo, A. Coc, E. Vangioni-Flam.
Atomic data and Nuclear data Tables. {\bf 88}, 203 (2004).
\bibitem{Lan86}
K. Langanke, Nucl. Phys. A {\bf 457}, 351 (1986).
\bibitem{Cs00}
A. Cs\'{o}t\'{o} and K. Langanke, Few-body Systems. {\bf 29}, 121
(2000).
\bibitem{Nol01}
 K.M. Nollett, Phys. Rev. C {\bf 63}, 054002 (2001).
 \bibitem{Dub95}
 S.B. Dubovitchenko, A.V. Dzhazairov - Kakhramanov, Yad. Fiz. {\bf 58}, 635 (1995)
[Phys.  At.  Nucl.  {\bf 58}, 579 (1995)]. {\bf 48}, 1664 (1982);
Nucl.Phys. A {\bf 419}, 115 (1984).
\bibitem{Moh93}
 P. Mohr, H. Abele, R. Zwiebel, G. Staudt, H. Krauss, H. Oberhummer,
  A. Denker, J.W. Hammer and F. Wolf, Phys. Rev. C {\bf 48}, 1420 (1993).
  \bibitem{Moh2009}
   P. Mohr, Phys. Rev. C {\bf 79}, 065804 (2009).
\bibitem{Igam97}
 S.B. Igamov, T.M. Tursunmuratov, and  R. Yarmukhamedov,Yad.Fiz. {\bf 60}, 1252 (1997)
[Phys.  At.  Nucl.  {\bf 60}, 1126 (1997)].
\bibitem{Osb82}
J.L. Osborne, C. A. Barnes, R.W. Ravanagh, R.M. Kremer, G.J.
Mathews, J.L. Zyskind, P.D. Parker, and A.J. Howard, Nucl.Phys.A
{\bf 419}, 115(1984).
\bibitem{Chris61}
R.F. Christy and I. Duck, Nucl.Phys. {\bf 24}, 89 (1961).
\bibitem{Igam07}
 S.B. Igamov, and R. Yarmukhamedov, Nucl.Phys.A {\bf 781}, 247 (2007);Nucl.Phys.A {\bf 832}, 346 (2010).
\bibitem{Blok77}
 L.D. Blokhintsev, I.Borbely, E.I. Dolinskii, Fiz.Elem.Chastits At. Yadra.
{\bf 8}, 1189 (1977)[Sov. J. Part. Nucl. {\bf 8}, 485 (1977)].
\bibitem{Blokh2008}
L.D. Blokhintsev and V.O. Yeromenko, Yad.Fiz. {\bf 71}, 1219 (2008)
[Phys. At.  Nucl.  {\bf 71}, 1126 (2008)].
\bibitem{Robert81}
R.G.H. Robertson, P. Dyer, R.A. Warner, R.C. Melin, T.J. Bowles,
A.B. McDonald, G.C. Ball, W.G. Davies, E.D. Earle, Phys.Rev.Lett.
{\bf 47},  1867 (1981).
\bibitem{Kim87}
K.H. Kim, M.H. Park, B.T. Kim, Phys.Rev. C {\bf 35}  363 (1987).
\bibitem{Igam08}
S.B. Igamov and R. Yarmukhamedov,  Phys. At.  Nucl.  {\bf 71}, 1740
(2008)].
\bibitem{Art2009}
S.V. Artemov, S.B. Igamov, K.I. Tursunmakhatov, and R.
Yarmukhamedov, Izv. RAN: Seriya Fizicheskaya, {\bf 73}, 176
(2009)[Bull.RAS:Physics,{\bf 73}, 165 (2009)].
\bibitem{Neu72}
 V.G. Neudatchin, V.I. Kukulin, A.N. Boyarkina and V.D. Korennoy, Lett.Nuovo
Cim.{\bf 5}, 834 (1972).
\bibitem{Neu75}
 V.I. Kukulin, V.G. Neudatchin, and Yu. F. Smirnov, Nucl.Phys. A {\bf 245}, 429 (1975).
\bibitem{Neu83}
 V.I. Kukulin, V.G. Neudatchin, I.T. Obukhovsky and Yu.F. Smirnov. "Clusters as
 Subsystems in Light Nuclei". In Clustering Phenomena in nuclei, Vol.3, eds.
 K. Wildermuth and P. Kramer ( Vieweg ,
 Braunschweig, 1983)p.1.
 \bibitem{Dub84} S.B.  Dubovichenko and M.A.
Zhusupov, Ivz.  Akad.  Nauk Kaz.SSR, Ser.  Fiz.-Mat. {\bf 4}, 44
(1983); Yad. Fiz. {\bf 39}, 1378 (1984)  [Sov. J.  Nucl. Phys.{\bf
39}, 870 (1984).]
\bibitem{Barn64}
A.L.C. Barnard, C. M. Jones, and G.C. Phillips, Nucl.Phys. {\bf 50},
629 (1964).
\bibitem{ST1967}
R. J. Spiger, and T. A. Tombrello, Phys.Rev. {\bf 163} (1967) 964.
\bibitem{Wall84}
H. Walliser, H. Kanada, and Y.C. Tang, Nucl.Phys. A {\bf 419}, 133
(1984).
\bibitem{Tim07}
N.K. Timofeyuk, P. Descouvemont, R.C. Johnson, Phys.Rev. C {\bf 75},
034302 (2007).
\bibitem{Cyb2008}
R. H. Cyburt and B. Davids, Phys.Rev.C {\bf 78}, 064614(2008).
\bibitem{Kurath75}
D. Kurath and D. J. Millener, Nucl.Phys.A {\bf 238}, 269 (1975).
\bibitem{Brune1999}
C.R. Brune, W.H. Geist, R.W. Kavanagh, K.D. Veal, Phys.Rev.Lett.
{\bf 83}, (1999) 4025.
\bibitem{Kra82}
H.Kr\"{a}winkel, H.W. Becker, L. Buchmann, J. G\"{o}rres, K.U.
Kettner, W.E. Kieser, R. Santo, P. Schmalbrock, H.P. Trautvetter, A.
Vlieks, C. Rolfs, J.W. Hammer, R.E. Azuma, W.S. Rodney, Z.Phys. A
{\bf 304}, 307 (1982).
\bibitem{Alt88}
T. Altmeyer, E. Kolbe, T. Warmann, K. Langanke, H.J. Assenbaum,
Z.Phys. A {\bf 330}, 277 (1988).
\bibitem{Sch87}
U. Schr\"{o}der, A. Redder, C. Rolfs, R.E. Azuma,  L. Buchmann, C.
Campbell, J.D. King, T.R. Donoghue, Phys.Lett. B {\bf 192}, 55
(1987).
\bibitem{Gon82}
S. A. Goncharov, J. Dobesh, E. I. Dolinskii, A. M. Mukhamedzhanov
and J.
 Cejpek, Yad. Fiz.{\bf 35}, 662 (1982)[Sov. J. Nucl. Phys. {\bf 35}, 383  (1982)].
 \bibitem{Mukh2005}
A. M. Mukhamedzanov, F. M. Nunes, Phys.Rev. C {\bf 72}, 017602
(2005).
\end{thebibliography}
\end{document}